\documentclass[notitlepage]{cernyrep}
\usepackage{texnames}

\usepackage{subfloat}
\usepackage[skip=6pt,font=small]{caption}    
\usepackage[skip=2pt,font=small,subrefformat=parens]{subfig}

\usepackage[T1]{fontenc}
\usepackage{url}
\usepackage{amsmath}
\usepackage{epstopdf}

\usepackage[labelfont=bf]{caption}
\usepackage{varwidth}
\usepackage{xcolor}

\newcommand{\beq}{\begin{equation}}
\newcommand{\eeq}{\end{equation}}

\newcommand{\eps}{\varepsilon}

\newcommand{\epsc}{\underline{\varepsilon}}
\newcommand{\lambdac}{\underline{\lambda}}
\newcommand{\muc}{\underline{\mu}}

\renewcommand{\rho}{\varrho}
\renewcommand{\theta}{\vartheta}
\renewcommand{\phi}{\varphi}

\newcommand{\ex}{\ensuremath{ \vec{e}_x} }

\newcommand{\ez}{\ensuremath{ \vec{e}_z} }

\newcommand{\erho}{\ensuremath{ \vec{e}_\varrho} }

\newcommand{\Wv}{\ensuremath{ \protect\vec{W} }}

\newcommand{\Ev}{\ensuremath{ \vec{E} }}
\newcommand{\Dv}{\ensuremath{ \vec{D} }}
\newcommand{\Hv}{\ensuremath{ \vec{H} }}
\newcommand{\Bv}{\ensuremath{ \vec{B} }}

\newcommand{\Jv}{\ensuremath{ \vec{J} }}

\newcommand{\Fv}{\ensuremath{ \vec{F} }}

\newcommand{\Zvc}{\ensuremath{\protect \underline{\vec{Z}} }}

\newcommand{\Evc}{\ensuremath{ \underline{\vec{E}} }}
\newcommand{\Dvc}{\ensuremath{ \underline{\vec{D}} }}
\newcommand{\Hvc}{\ensuremath{ \underline{\vec{H}} }}
\newcommand{\Bvc}{\ensuremath{ \underline{\vec{B}} }}

\newcommand{\Jvc}{\ensuremath{ \underline{\vec{J}} }}
\newcommand{\Avc}{\ensuremath{ \underline{\vec{A}} }}

\newcommand{\Rvc}{\ensuremath{ \underline{\vec{R}} }}
\newcommand{\Fvc}{\ensuremath{ \underline{\vec{F}} }}

\newcommand{\Zc}{\ensuremath{\protect \underline{Z} }}
\newcommand{\Uc}{\ensuremath{ \protect\underline{U} }}

\newcommand{\Gc}{\ensuremath{ \underline{G} }}
\newcommand{\Ec}{\ensuremath{ \underline{E} }}
\newcommand{\Ic}{\ensuremath{ \underline{I} }}

\newcommand{\Hc}{\ensuremath{ \underline{H} }}
\newcommand{\Fc}{\ensuremath{ \underline{F} }}

\newcommand{\Jc}{\ensuremath{ \underline{J} }}

\newcommand{\Phic}{\ensuremath{ \underline{\Phi} }}

\newcommand{\rhoc}{\ensuremath{ \underline{\rho} }}
\newcommand{\nuc}{\ensuremath{ \underline{\nu} }}

\newcommand{\rv}{\ensuremath{ \vec{r} }}
\newcommand{\dv}{\ensuremath{ \vec{d} }}
\newcommand{\pv}{\ensuremath{ \vec{p} }}

\newcommand{\nv}{\ensuremath{ \vec{n} }}

\newcommand{\vv}{\ensuremath{ \vec{v} }}

\newcommand{\intinf}{\int_{-\infty} ^\infty }

\newcommand{\dsv}{\ensuremath{ \mathrm{d}\vec{s}} }

\newcommand{\dAv}{\ensuremath{ \mathrm{d}\vec{A}} }
\newcommand{\ds}{\ensuremath{ \mathrm{d}s}}

\newcommand{\dV}{\ensuremath{ \mathrm{d}V}}
\newcommand{\dx}{\ensuremath{ \mathrm{d}x}}

\newcommand{\dz}{\ensuremath{ \mathrm{d}z}}
\newcommand{\dt}{\ensuremath{ \mathrm{d}t}}

\newcommand{\domega}{\ensuremath{ \mathrm{d}\omega}}

\newcommand{\ul}{\underline}

\newcommand{\wegdamit}[1]{} 

\newlength{\lwveryfine}   
\newlength{\lwfine}   
\newlength{\lwnormal} 
\newlength{\lwthick}  
\newlength{\lwverythick}  

\usepackage{FIT}

\begin{document}
\title{Bench Measurements and Simulations of Beam Coupling Impedance}

\author{U. Niedermayer}

\institute{Institut f\"ur Theorie Elektromagnetischer Felder, Technische Universit\"at Darmstadt, Germany}

\begin{abstract}
After a general introduction, the basic principles of wake-field and beam-coupling-impedance computations are explained.
This includes time domain, frequency domain, and methods that do not include excitations by means of a particle beam.
The second part of this paper deals with radio frequency bench measurements of beam coupling impedances. The general procedure of the wire measurement is explained, and its features and limitations are discussed.
\end{abstract}


\maketitle 

\section{Introduction: Maxwell's equations, wakes and impedance}
A complete macroscopic description of electromagnetic (EM) fields as function of position $\rv\in\Omega\subset\mathbb R^3$ and time $t\in\mathbb R$ is given by Maxwell's equations
\begin{subequations}
\label{MaxwellTD}
	\begin{align}
		\label{faraday}
				\nabla \times \Ev (\rv,t) &= -\partial_t \Bv(\rv,t) \\
		\label{ampere}
			\nabla \times \Hv(\rv,t) &=\Jv_{s}(\rv,t) + \Jv(\rv,t) + \partial_t \Dv(\rv,t) \\
		\nabla \cdot \Dv(\rv,t) &=  \rho_s(\rv,t)  \\
		\nabla \cdot \Bv(\rv,t) &= 0
	\end{align}
\end{subequations}
and material equations
\begin{subequations}
\label{MaterialTD}
	\begin{align}
   \Dv(\rv,t) &= \varepsilon(\rv) \Ev(\rv,t) \\
   \Bv(\rv,t) &= \mu(\rv) \Hv(\rv,t) \\
   \Jv(\rv,t) &= \kappa(\rv) \Ev(\rv,t)~,
	\end{align}
\end{subequations}
where $\rho_s$ and $\Jv_s$ denote the source charge and current densities, respectively.
The material distribution is given by the permittivity $\eps$, the conductivity $\kappa$ and the permeability $\mu$, which are assumed to be linear and isotropic.
At first, the material is also assumed to be non-dispersive. By means of Gauss' and Stokes' theorems, Maxwell's equations can also be written in integral form, which is more general, since the differentiability requirements of the field vector functions can be relaxed.

The force acting on a point-like charged particle $i=1,\ldots, N$ is
\begin{align}
\label{LorentzForce}
\Fv_i(t)=q\left(\Ev(\rv_i,t)+\vv_i(t)\times\Bv(\rv_i,t)\right),
\end{align}
where $q$ and $\vv_i$ are the particle's charge and velocity, respectively.
\begin{figure}[h!]
\subfloat[Description of particle motion in EM-fields. For full analytic self-consistency $\Delta t\rightarrow 0$ is required. Numerical approaches can, dependent on the time scale of the problem, allow finite $\Delta t$ and still be self-consistent.]{
		\includegraphics[width=0.46\textwidth]{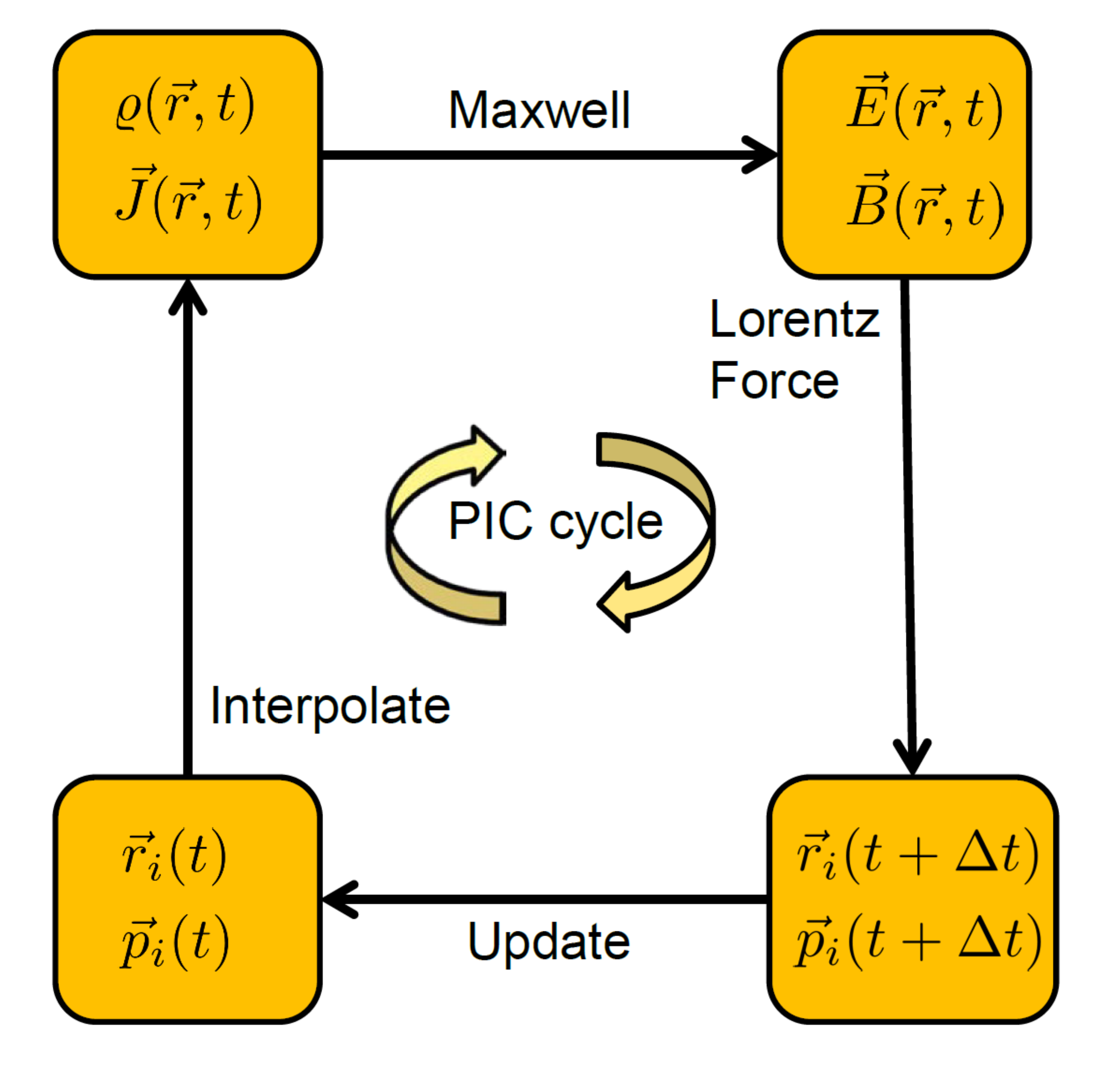}
			\label{fig:PICa}
		}	
		\hfill
		\subfloat[Description of particle motion with wake fields. The EM forces are precomputed using a rigid beam as source. For synchrotrons, the time step is usually one revolution period, which is a good approximation for multi-turn phenomena.
		]{
		\includegraphics[width=0.46\textwidth]{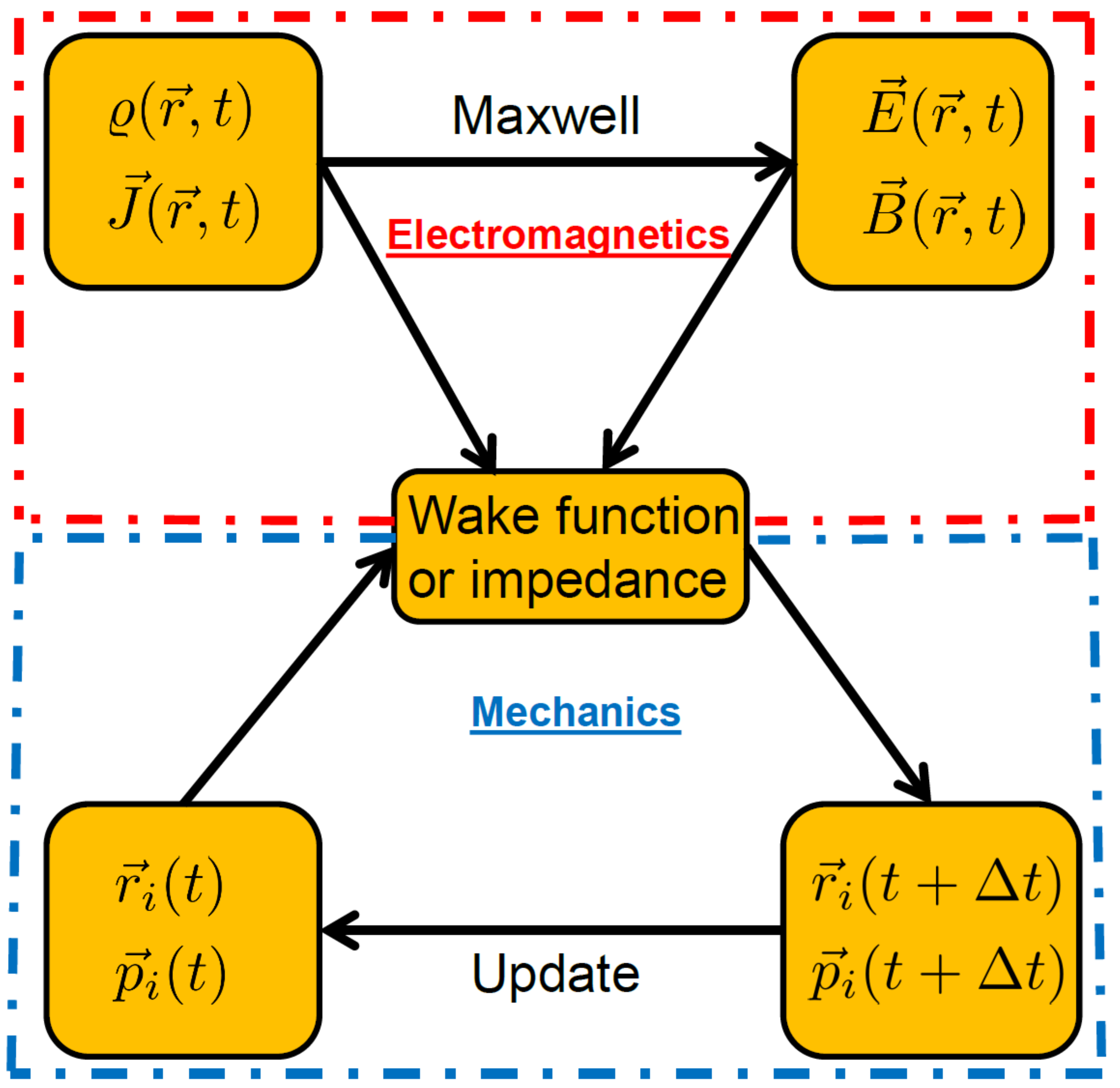}
			\label{fig:PICb}
		}

		\caption[PIC loop vs wake-field approach.]{PIC simulation loop versus wake-field approach}
	\label{fig:PIC}
\end{figure}
For a description of particle motion in both their own and external fields, Maxwell's equations have to be coupled with the equations of mechanics,
\begin{align}
\label{ForceDef}
\Fv_i (t)&=\partial_t \pv_i(t) \\
\label{MomentumDef}
\pv_i(t)&=\gamma_i(t) m_i \vv_i(t)~,
\end{align}
where the relativistic mass and velocity factors are $\gamma=(1-\beta^2)^{-1/2}$ and $\beta=|\vec{v}|/c$,
and $\pv_i(t)$ is the $i$th particle's momentum.
In order to simulate the dynamics of many particles self-consistently, mostly particle in cell (PIC) or Vlasov--Maxwell solvers are used (cf{.} Fig{.}~\ref{fig:PIC}(a)).
The wake function approach is different, since Maxwell's equations are entirely decoupled from the equations of motion (cf{.} Fig{.}~\ref{fig:PIC}(b)).
Therefore one obtains self-consistency only for phenomena which evolve slower than the wake-field kicks.

\subsection{Wake functions}
Usually one makes two assumptions that decouple mechanics from electromagnetics (cf{.} Fig{.}~\ref{fig:PIC}(b)).
\begin{enumerate}
\item \textbf {Rigid Beam Approximation} Although the leading charge loses energy, its velocity remains unchanged.
																															 This is exactly fulfilled for an ultrarelativistic beam which carries infinite energy.
\item \textbf {Kick Approximation}		The wake force continuously acting on the trailing charge is lumped in a single kick \textit{after} the passage through the device.
																															 This means that the trailing charge is also assumed to be rigid during the passage.
\end{enumerate}
These approximations are justified by the different time scales of the particle passage (fast) and the evolution of wake-field effects (slow).

\begin{figure}[h]
  \centering
	\includegraphics[width=0.8\textwidth]{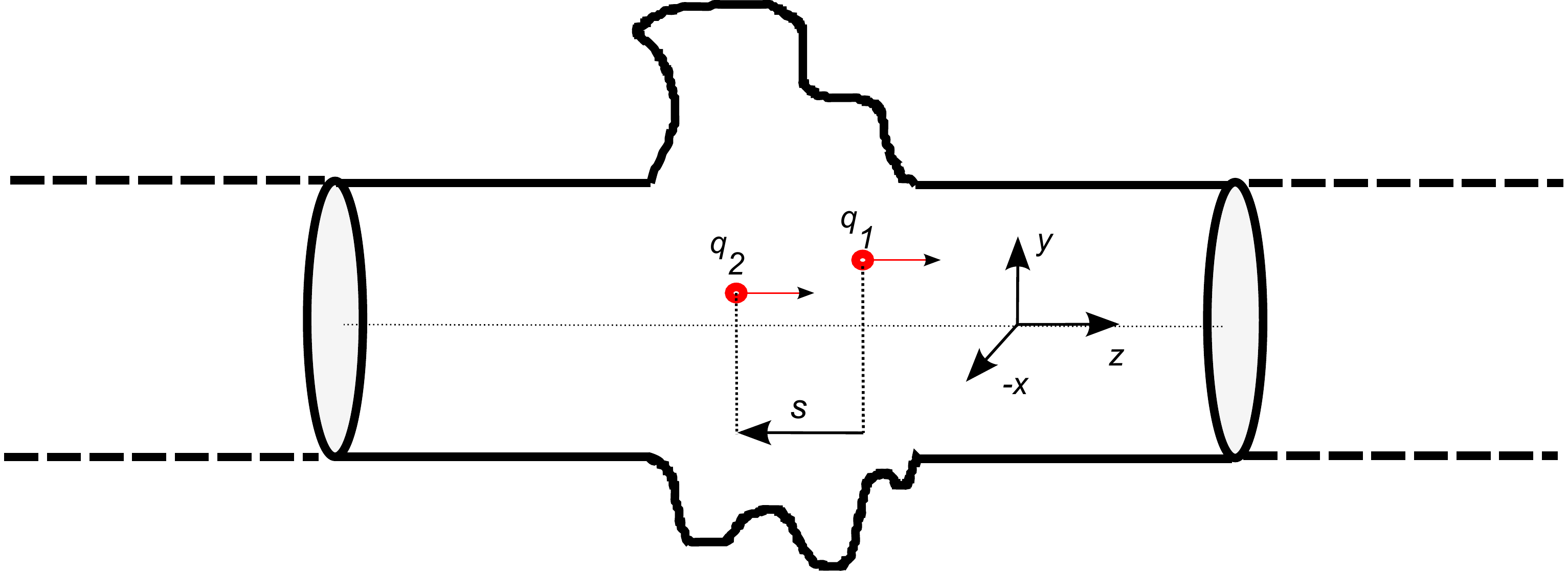}
	\caption[Definition of the wake function.]{Longitudinal cut through an accelerator structure and centred coordinate system. The charge $q_1$ is usually referred to as source or leading charge and $q_2$ is referred to as test or trailing charge.}
	\label{fig:wakedef}
\end{figure}

We define the wake function as (see also, for example, Ref{.}~\cite{Weiland1993,Palumbo1994})
\begin{align}
  \Wv(\rv_2^\perp,\rv_1^\perp,s):
  &=
  \frac{1}{q_1 q_2}\intinf
  \Fv\left(\rv_2,z_2,t=\frac{z_2+s}{v}\right) \,\dz_2 \nonumber\\
  &=
  \frac{1}{q_1}\intinf \left[\Ev+\vv \times \Bv\right] \left(\rv_2,z_2,t=\frac{z_2+s}{v}\right)   \,\dz_2~,
  \label{wakefunction}
\end{align}
such that a positive value indicates momentum or energy gain for the test charge.
The integral in Eq{.}~(\ref{wakefunction}) exists only if the assumed infinitely long pipe connections (see Fig{.} \ref{fig:wakedef}) do not cause any wake fields, which requires the following conditions:
\begin{itemize}
\item smooth pipe (no geometric wake fields);
\item perfectly conducting pipe (no resistive wake fields);
\item ultrarelativistic beam (no space-charge wake fields).
\end{itemize}


When all these conditions are fulfilled, the infinite integration in Eq{.}~(\ref{wakefunction}) can be replaced by a finite one, since
the scattered fields from the 3D region decay in the pipe below the waveguide cutoff frequency.
Above the cutoff frequency, waveguide modes do not interact with the particle beam in average, since the fields are periodic and
$v_\mathrm{beam}\leq c<v_\mathrm{phase}^\mathrm{mode}$.
An estimate of the decay length for a given threshold can be found in Ref{.}~\cite{Balk}.
For non-ultrarelativistic velocity, sophisticated boundary conditions are required for the entry and exit of the beam in a finite-sized computational domain, see \cite{Balk2006}.
In order to ensure that the integral in Eq{.}~(\ref{wakefunction}) is finite, even if the third assumption is not fulfilled, the integration can be performed over a finite length and the space-charge interaction
in the infinitely long pipe can be described by a space-charge wake function per length.
Note that the three components of Eq{.}~(\ref{wakefunction}) are not independent, but connected by the Panofsky--Wenzel theorem~\cite{Panofski1956}, which can be conveniently expressed as $\nabla'\times \Wv(\rv_2^\perp,s)=0$, where the curl acts on the relative position of the trailing charge, i.e{.} $\nabla'=(\partial_{x_2},\partial_{y_2},-\partial_{s})^T$.
The longitudinal wake function reads
\beq
W_\parallel(s)= \frac{1}{q_1 q_2}\intinf  \Ev \left(\rv_2=0,z_2,t=\frac{z_2+s}{v}\right)   \,\dz_2
\eeq
and is directly connected with the energy loss of the trailing charge by $\Delta E(s)=-q^2 W_\parallel(s)$.
The energy loss of a test charge caused not only by a single charge but by a whole bunch can be found by convolution with the beam distribution as
\beq
W^\mathrm{pot}_\parallel(s)=\intinf W_\parallel(s') \lambda (s-s')\, \ds'
\eeq
and is called the wake potential. If $\lambda$ is normalized to 1, the energy loss of a test charge becomes $\Delta E(s)=-N q^2 W^\mathrm{pot}_\parallel(s)$, where $N$ is the number of particles in the bunch.

In order to quantify the transverse deflection of the beam by wake fields one makes the assumption of small displacements from the beam axis ($d_{x/y_{1/2}}$ of the leading and trailing charge in the $x/y$ direction, respectively) and defines
\beq
W_{\perp,x}^\mathrm{drive}(s)=\frac{1}{q_1 d_{x_1}}\intinf \left[\Ev+\vv \times \Bv\right] \left(\rv_2^\perp=0,z,\frac{z_2+s}{v}\right)   \,\dz_2
\eeq
\beq
W_{\perp,x}^\mathrm{det}(s)=\frac{1}{q_1 d_{x_2}}\intinf \left[\Ev+\vv \times \Bv\right] \left(\rv_2^\perp=d_{x_2}\ex,z,\frac{z_2+s}{v}\right)   \,\dz_2
\eeq
in units of $\mathrm{VA}^{-1}\mathrm{s}^{-1}\mathrm{m}^{-1})$. The linearized transverse kick is thus $\Delta p_x(s)=q^2/\beta c  (d_{x_1}W_{\perp,x}^\mathrm{drive}(s)+d_{x_2}W_{\perp,x}^\mathrm{det}(s))$.
The dipolar or driving wake acts coherently, since the force experienced by a test charge does not depend on the test charge's position in a dipole field.
Contrarily, the detuning wake acts incoherently, as it depends on the displacement of the test charge linearly.

\begin{figure}[t]
  \centering
		\includegraphics[width=0.4\textwidth]{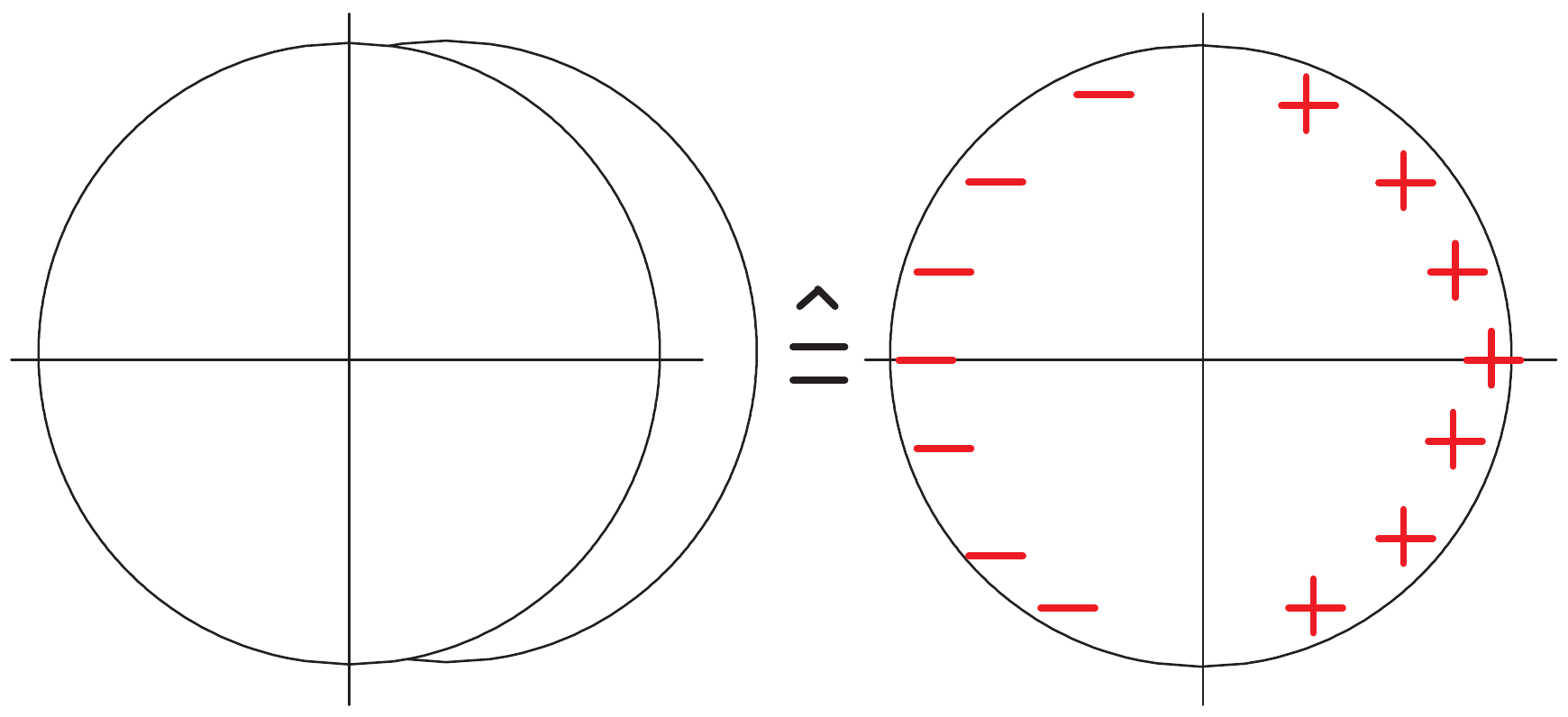}
	\caption{Illustration of a coherent transverse oscillation (left-hand side) represented by a dipole moment (right-hand side).}
	\label{fig:dipolemode}
\end{figure}

If the beam is not ultrarelativistic, space-charge effects need to be taken into account and the transverse beam dimensions need to be considered as finite.
We assume the beam to be transversely uniform of radius $a$. A coherent displacement can be written as
\beq
\sigma(\rv^\perp)=\frac{1}{\pi a^2}\Theta(a-|\rv^\perp-\dv^\perp|)~,
\label{sigmadisplaced}
\eeq
where $\dv^\perp$ is the transverse displacement vector and $\Theta$ denotes the Heaviside step function.
This displacement can be approximated in polar coordinates $\rho,\phi$ by (Ref{.}~\cite{Khateeb2007})
\beq
 \sigma(\rho,\phi)\approx \frac{1}{\pi a^2}( \Theta(a-\rho)+\delta(a-\rho)(d_\mathrm{x}\cos{\phi}+d_\mathrm{y}\sin{\phi})+\cdots )
 \label{sigma}
\eeq
which in first order is a dipole moment, see Fig{.}~\ref{fig:dipolemode}.

\subsection{Frequency domain}
In this framework, the following convention for the Fourier transform is used:
\beq
\Fc(\omega)=\mathcal{F}\{f(t)\}(\omega)=\intinf f(t) \text{e}^{-\text{i}\omega t} \,\dt.
\label{Eq:fourier}
\eeq
The inverse transform reads
\beq
f(t)=\mathcal{F}^{-1}\{\Fc(\omega)\}(t)=\frac{1}{2\pi}\intinf \Fc(\omega) \text{e}^{+\text{i}\omega t} \,\domega~,
\label{Eq:invfourier}
\eeq
and between the two domains Plancherel's theorem holds, $2\pi(f,g)=(\Gc,\Fc)$, where the brackets denote a scalar product $(f,g)=\int f g^* \,\dx$ and $^*$ is complex conjugation.
The duration and bandwidth of a mean value free signal $f(t)$ are defined with $\|\cdot\|_2=\sqrt{(\cdot,\cdot)}$ as (see  Ref.~\cite{Unbehauen})
\beq
T=\frac{\|tf(t)\|_2}{\|f(t)\|_2},  \;\;\;\;
B=\frac{\|\omega\Fc(\omega)\|_2}{\|\Fc(\omega)\|_2}~.
\eeq
The two fulfil the K\"upfm\"uller uncertainty principle, see Ref{.}~\cite{KM}, $T\cdot B \geq 1/2$,
where equality is achieved for a Gaussian pulse.
Applying Eq{.}~\eqref{Eq:fourier} to Eqs.~\eqref{MaxwellTD}, Maxwell's equations read in the frequency domain
\begin{subequations}
	\label{maxwellF}
	\begin{align}
		\label{faradayF}
			\nabla \times \Evc (\rv,\omega) &= -\text{i}\omega \Bvc(\rv,\omega) \\
		\label{ampereF}
			\nabla \times \Hvc(\rv,\omega) &=\Jvc_{s}(\rv,\omega) + \Jvc(\rv,\omega) + \text{i}\omega \Dvc(\rv,\omega) \\
		\nabla \cdot \Dvc(\rv,\omega) &=  \rhoc_s(\rv,\omega)  \\
		\nabla \cdot \Bvc(\rv,\omega) &= 0~,
	\end{align}
\end{subequations}
where the material relations can easily include dispersive materials as
\begin{subequations}
	\begin{align}
   \Dvc(\rv,\omega) &= \eps_0\epsc_r(\rv,\omega) \Evc(\rv,\omega) \\
   \Bvc(\rv,\omega) &= \mu_0\muc_r(\rv,\omega) \Hvc(\rv,\omega) \\
   \Jvc(\rv,\omega) &= \kappa(\rv,\omega) \Evc(\rv,\omega)~.
	\end{align}
\end{subequations}
Combining Eqs.~\eqref{maxwellF} leads to the curl--curl equation
\beq
\nabla\times\muc^{-1}\nabla\times\Evc+ \text{i}\omega\kappa \Evc-\omega^2\eps \Evc =-\text{i} \omega \Jvc_{s}
\label{curlcurl}
\eeq
and the continuity equation
\beq
\nabla \cdot (\Jvc_s+\kappa\Evc)+\text{i}\omega \rhoc_s=0~.
\eeq
The beam current density is modelled as a convection current density $\Jvc_s=\rhoc_s v \ez$.
Therefore, the spatial Fourier correspondence $\partial_z \rightarrow -\text{i}k_z$ is given for the source fields (beam in free space) by
$k_z=\omega/v$.
In a longitudinally homogeneous and smooth two-dimensional (2D) structure, this property must hold also for the fields scattered back from the wall.

The beam coupling impedance is defined as the Fourier transform of the wake function
\beq
\Zvc(\rv_1^\perp,\rv_2^\perp,\omega)=-\intinf \Wv(\rv_1^\perp,\rv_2^\perp,s) \text{e}^{-\text{i}\omega s/v} \,\frac{\ds}{v}
=-\frac{1}{q_1 q_2}\int_{-\infty}^{\infty} \vec\Fc(\rv_1^\perp,\rv_2^\perp,z,\omega) \text{e}^{+\text{i}\omega z/v} \,\dz~,
\label{impedancedef}
\eeq
where $\Fvc(\omega)=q(\Evc(\omega)+\vv\times\Bvc(\omega))$ is the spectral density of the force. If the beam has a finite transverse size, the longitudinal impedance can be written as (see Ref{.}~\cite{Gluckstern2000})
\beq
\label{ZlVol}
\Zc_\parallel(\omega)=-\frac{1}{q^2} \int_\mathrm{beam} \Evc \cdot \Jvc_s^* \,\dV~.
\eeq
The transverse impedance is usually defined with an extra $(-\text{i})$, in order to relate the real part to instability growth and the imaginary part to phase shift, as it is the case for the longitudinal impedance.
The dipolar transverse impedance reads
\beq
\Zc^\mathrm{drive}_{\perp,x/y}(\omega,\rv_2)=-\frac{(-\text{i})}{q_1 d_{x_1/y_1}}  \int_{-\infty}^{\infty}\left[\Evc(\omega)+\vv \times \Bvc(\omega) \right]_{x/y} \text{e}^{\text{i}\omega z/v} \,\dz~,
\label{Ztdef}
\eeq
which can be generalized using the Panofsky--Wenzel theorem to (see Ref{.}~\cite{Gluckstern2000})
\beq
\label{ZtVol}
\Zc^\mathrm{drive}_{\perp,x/y}(\omega)=-\frac{v}{\omega(q d_{x/y})^2} \int_\mathrm{beam} \Evc \cdot \Jvc_{s, d_{x/y}}^* \,\dV~,
\eeq
where $\Jvc_{s, d_{x/y}}$ is the dipolar component of the source current density in FD, cf{.} Eq{.}~\eqref{sigma}.
Note that the formulations of Eqs.~\eqref{ZlVol} and \eqref{ZtVol} are particularly convenient for evaluation on a mesh, since evaluation errors are averaged out by the integration.

\subsection{Helmholtz decomposition}
\label{Helmholtz}
In a simply connected domain $\Omega\subset\mathbb{R}^3,$ any differentiable vector field
$\Avc: \Omega \rightarrow \mathbb{C}^3$
can be written as $\Avc=\Avc_\mathrm{curl}+\Avc_\mathrm{div}$, where $\Avc_\mathrm{curl}$ and $\Avc_\mathrm{div}$ are uniquely determined by demanding $\nabla \times \Avc_\mathrm{div}=0$ and   $\nabla\cdot \epsc\, \Avc_\mathrm{curl}=0$ for a piecewise smooth non-vanishing function $\epsc: \Omega \rightarrow \mathbb{C}$.
See, for example, Ref.~[p.~86]\cite{Monk} for a proof.

If the domain $\Omega$ is not simply connected, the Helmholtz decomposition has to be generalized to the so-called Hodge decomposition, i.e{.} a third field becomes constituent of $\Avc$.
This so-called harmonic field satisfies both  $\nabla \times \Avc_\mathrm{harm}=0$ and $\nabla\cdot \epsc\,\Avc_\mathrm{harm}=0$ and is yet non-zero.


Applying the Helmholtz decomposition with $\Evc_\mathrm{div}=-\nabla\Phic$ to Eq{.}~\eqref{curlcurl} we find
\begin{subequations}
\label{CurlcurlHelmholzSplit}
\begin{align}
\label{Poisson}
-\nabla\cdot \epsc \nabla \Phic&= \rhoc_s \\
\label{CurlCurlDivFree}
\nabla\times\nuc\nabla\times\Evc_\mathrm{curl}-\omega^2\epsc \, \Evc_\mathrm{curl} &=\Rvc~,
\end{align}
\end{subequations}
where $\rhoc_s=(\mathrm{i}/\omega)\nabla\cdot\Jvc_{s}$ and
$
\Rvc=-\omega^2\epsc \nabla \Phic - \mathrm{i} \omega \Jvc_{s}.
$
It is crucial here that due to the continuity equation $\nabla\cdot\Rvc=0$ holds, i.e{.} all vector fields in Eq{.}~\eqref{CurlCurlDivFree} are divergence free.
Moreover, for a beam in $z$-direction, the charge can be written as $\rhoc_s=\Jc_{s,z}/v$.

\section{Impedance simulations}
Beam coupling impedance simulations can be sorted into three main groups, namely Time Domain (TD), Frequency Domain (FD), and methods without a particle beam.
Most common are explicit TD methods, since they require only matrix-vector multiplications for time stepping.
They are usually based on finite differences time domain (FDTD, Yee~1966~\cite{Yee1966}) or finite integration technique (FIT, Weiland 1977~\cite{Weiland1977}), which result in a coinciding space discretization on a Cartesian mesh. However, note that, in general, mesh and method are independent, e.g{.} FIT or the finite element method (FEM) can be applied on both structured and unstructured, tetrahedral or hexahedral, or even mixed meshes.



\begin{table}[h]
\caption{Examples of time domain wake-field codes. More detailed summaries can be found e.g{.} in \cite{GjonajICFA, Handbook}}
	\centering
\begin{tabular}{lccc}
\\ \hline\hline
					Code				&   Method					&	   Website	  	 										         & Availability 		 \\ \hline
	CST PS &   FIT							&	\url{www.cst.de}												 & commercial  		\\
	GdfidL  						&   FDTD						&	 \url{www.gdfidl.de}													 & commercial  		 \\
		 		Echo					&   FIT							&	 \url{www.desy.de/fel-beam/s2e/codes.html}    & free  					\\
		PBCI 							&  	FIT, DG-FEM			&	 \url{www.temf.de} , see also \cite{GjonajICFA}                                        & on request  					 \\
		ABCI 							&  	FIT 						&	 \url{abci.kek.jp/abci.htm}                                         & free  					 \\
	  ACE3P							&  	implicit FEM 		& \url{slac.stanford.edu} & free in USA \\
\hline\hline
\end{tabular}
\label{Codes}
\end{table}

Explicit TD simulations are suitable at medium and high frequency, and particularly in perfectly conducting structures.
They are disadvantageous for low frequencies and low velocity of the beam. Also dispersively lossy materials are difficult to treat in TD, since a convolution with the impulse response, i.e{.} the inverse FT of the material dispersion curve, is required.
In FD the beam velocity and dispersive material data are just parameters. However, a system of linear equations (SLE) has to be solved for each frequency point, which is costly when the matrix is large and ill-conditioned.

In the following we focus on the FIT in TD and both FIT and FEM in FD.
More specialized techniques in TD are the boundary element method (BEM)~\cite{Kawaguchi1995,Fujita2006}, the finite volume (FV) method ~\cite{Gjonaj2009,Tsakanian2015}, discontinuous Galerkin finite element method (DG-FEM), and implicit methods.
In FD, there is no direct advantage from diagonal material matrices, which favours the FEM on an unstructured mesh. However, FIT is also used, since the structured mesh makes the implementation of Floquet boundary conditions simple. Particularly in the absence of materials, BEM is also an attractive option in FD~\cite{Yokoya1993,Macridin2013}.
An overview of some commonly used TD codes is given in Table~\ref{Codes}.

The most common method without particle beam is the computation of eigenmodes, which can be related to the wake function as discussed in Ref{.}~\cite{Bane1984}.
Eigenmode computations with FEM for perfectly electric conducting (PEC) structures including losses through beam pipes and couplers are described in Ref{.}~\cite{Ackermann2012}, a FIT algorithm for dispersively lossy tensorial materials used in e.g{.} ferrite cavities is presented in Ref{.}~\cite{Klopfer2015}.
Methods with excitations other than particle beams are based on current path~\cite{Kroyer2008,Niedermayer2012} or plane wave~\cite{Kononenko2011} excitation, but they require special interpretations to obtain the beam coupling impedance.

%

\subsection{(Explicit) time domain}
Due to the minimal duration--bandwidth product, the excitation is usually a Gaussian bunch
\beq
\lambda(z,t)=\frac{q}{\sqrt{2\pi}\sigma_s}\text{e}^{-\frac12\left(\frac{z-vt}{\sigma_s}\right)^2}~,
\eeq
which rigidly moves through the structure. The spectrum of this pulse is obtained from the FT over $s=vt-z$ as
\beq
\left|\lambdac(\omega)\right|=\frac{q_\omega}{\sqrt{2\pi}\sigma_\omega}\text{e}^{-\frac12\left(\frac{\omega}{\sigma_\omega}\right)^2}~,
\eeq
where $\sigma_\omega=v/\sigma_s$ and the normalization is $q_\omega=q\sqrt{2\pi}\sigma_\omega/v$.
The duration and bandwidth are
\beq
T=\frac{\sigma_s}{\sqrt2 v}, \;\;\;\;\; \;
B=\frac{v}{\sigma_s \sqrt2}~,
\eeq
resulting in $TB=1/2$. The point charge impedance is obtained by the convolution theorem as
\beq
\Zvc(\omega)=\frac{\mathcal{F}\{\Wv^\mathrm{pot}(s)\}(\omega)}{\mathcal{F}\{\lambda(s)\}(\omega)}~,
\eeq
where the numerically obtained wake potential is transformed by (equidistant) discrete Fourier transform (DFT).
The choice of the bunch length $\sigma_s$ does not necessarily depend on the real bunch length in the accelerator, but rather on the frequency of interest.
The maximum frequency at which a reasonable excitation amplitude is present, is roughly $2\sigma_f$, i.e{.} the spectrum is mainly located at $\sigma_f=v/(2\pi \sigma_s)$, the so-called frequency associated with the bunch length. Shorter bunches increase the maximum frequency, but they decrease the frequency resolution at low frequency.
The frequency resolution depends on the total number of points employed for an equidistant DFT as $\Delta f= \frac{1}{N_\mathrm{DFT}\Delta t}$ and
the total integrated wake length is $L_\mathrm{W}=v N_\mathrm{DFT}\Delta t$. Bunch length and wake length are the parameters to be set for the impedance simulation.

\begin{figure}[t]
	\centering
		\includegraphics[width=0.9\textwidth]{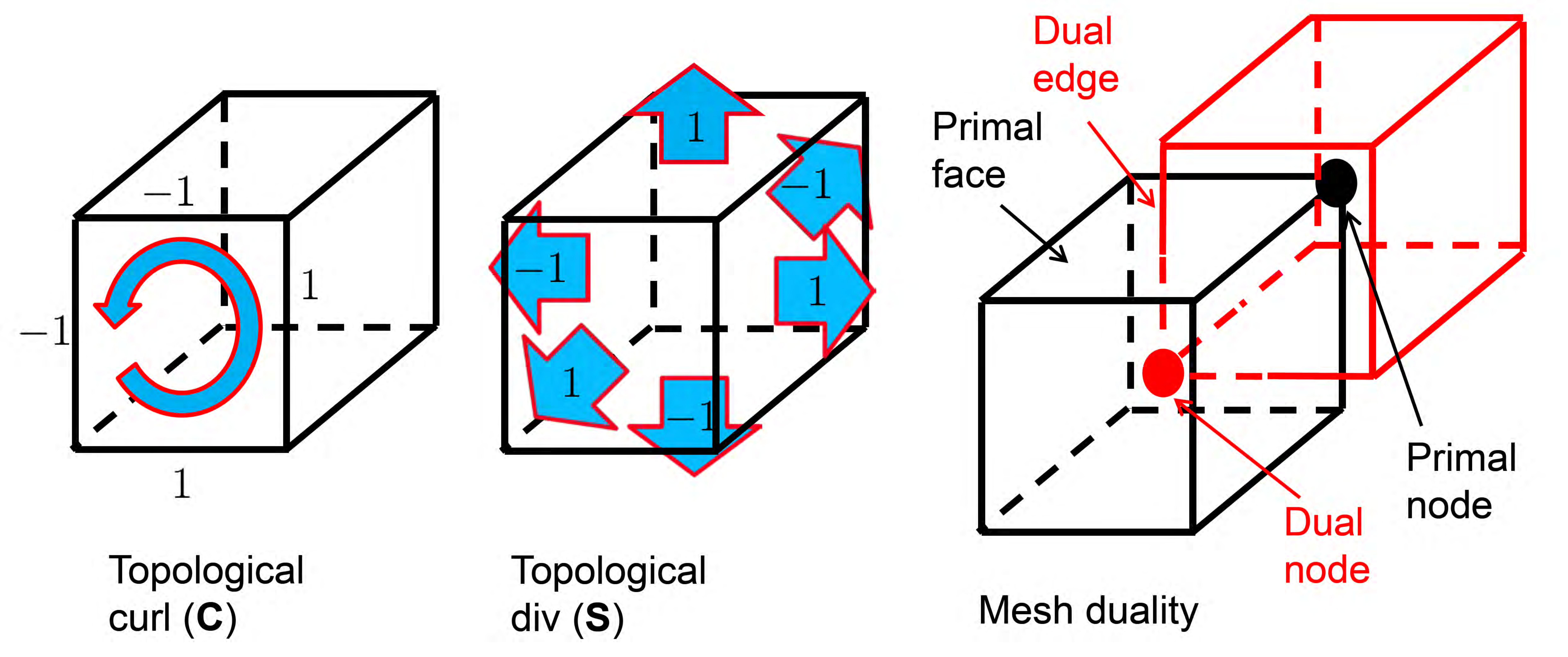}
		\caption{Topological FIT mesh properties}
	\label{FITpics}
\end{figure}

The FIT is based on evaluating the integral form of Maxwell's equations on a given mesh\footnote{For simplicity, we restrict ourselves here to Cartesian meshes.}, i.e{.}
\begin{align}
	&\ve_i = \int_{L_i} \Ev \cdot \dsv          	&\vh_i = \int_{\tilde L_i} \Hv \cdot \dsv  	&& \q_i = \int_{\tilde V_i} \rho \,\dV  \nonumber \\
	&\fd_i = \int_{\tilde A_i} \Dv \cdot \dAv 	  &\fb_i = \int_{A_i} \Bv \cdot\dAv 					 && \fj_i = \int_{\tilde A_i} \Jv \cdot \dAv~.
	\label{FITstate}
\end{align}
The resulting $3N_p$-dimensional vectors are the electric and magnetic edge voltages $\ve$ and $\vh$, face fluxes $\fd$ and $\fb$, and the face current $\fj$ and volume charge $\q$, which are connected by the continuity equation. Combining the integrals of Eqs.~\eqref{FITstate} to closed loops or closed surfaces (see Fig{.}~\ref{FITpics})
results in the Maxwell-Grid-Equations (MGE)
\begin{subequations}
	\label{MGE}
	\begin{align}
		\label{discretefaraday}
		\C\ve&=-\partial_t \fb \\
		\label{discreteampere}
		\Ct\vh&=\fj_s+\fj+ \partial_t \fd \\
		\label{discreteecharge}
		\St\fd&=\q_s \\
		\label{discretemcharge}
		\S\fb&=0~,
	\end{align}
\end{subequations}
where $\C$ and $\S$ are purely topological curl and divergence operators.
The operators $\Ct, \St$ in Eqs.~\eqref{MGE} represent evaluation on a dual grid, which has the property that dual vertices and edges intersect primal volumes and faces with same index, respectively, and vice versa.
The MGE Eqs.~\eqref{MGE} are exact, since they represent an evaluation of Maxwell's equations on the given grid topology. 	
The numerical approximations required to solve the MGE are included in the material matrices ($\nu=\mu^{-1}$)
\beq
		\label{MaterialMatrices}
		\vh=\fMnu\fb~, \\
\;\;\;\;\;\;\;\;\;\;\;\;
		\fd=\fMeps\ve~,
\;\;\;\;\;\;\;\;\;\;\;\;
		\fj=\fMkap\ve~,
\eeq	
which are diagonal matrices due to the dual orthogonal mesh and given by
	\begin{align}
			\left[\fMnu\right]_{n,n}&=\overline \nu_n \frac{\tilde L_n}{A_n} \approx \frac{\int_{\tilde L_n} \Hv \cdot \,\dsv}{\int_{A_n} \Bv \cdot\,\dAv }  \\
				\left[\fMeps\right]_{n,n}&=\overline \eps_n \frac{\tilde  A_n}{L_n} \approx\frac{\int_{\tilde A_n} \Dv \cdot\,\dAv }{\int_{L_n} \Ev \cdot \,\dsv} ,
				\label{MaterialApprox}
	\end{align}
where $L_n, \tilde L_n, A_n, \tilde A_n$ denote the length and area of the $n$th primal and dual cell edge and face, respectively.
	
The discretization of time derivatives can be done by forward (explicit) or backward (implicit) finite differences.
Implicit methods are unconditionally stable, but they require solving a SLE in each time step.
\begin{figure}[h!]
	\centering
   \includegraphics[width=0.5\textwidth]{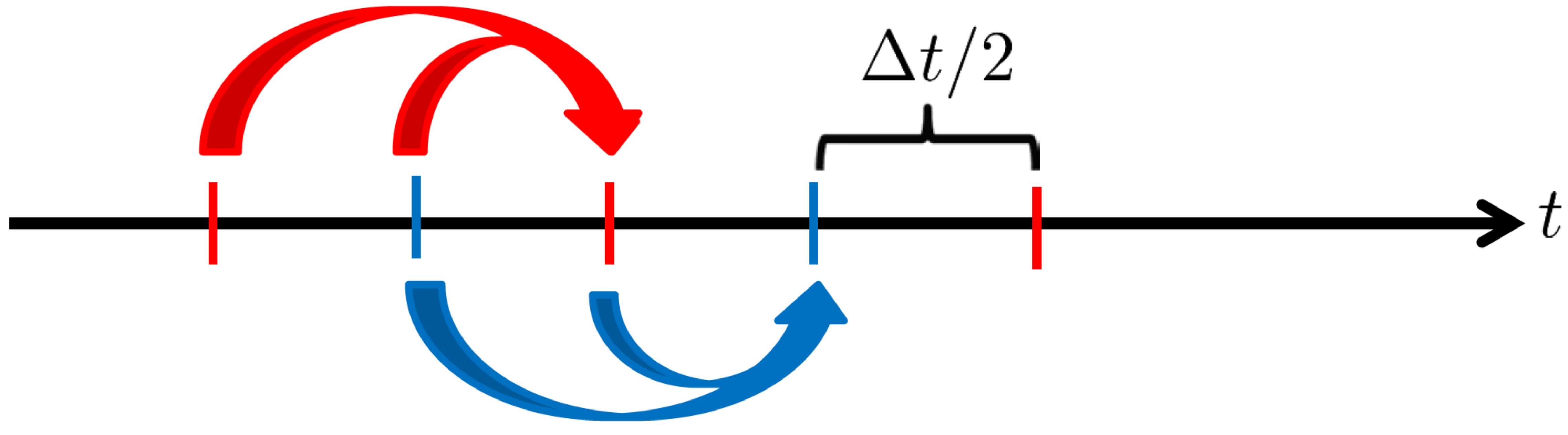}
    		\caption{Leap-frog method. Magnetic fields are allocated on half integer time indices between two electric field steps.}
    		\label{Leapfrog}
\end{figure}
Explicit methods are much lighter in computation, but they are only conditionally stable.
The most commonly used explicit method is the so-called `leap-frog' method, introduced by Yee~\cite{Yee1966} in 1966. It consists of a (staggered) central-difference quotient featuring second-order accuracy
\begin{subequations}
\begin{align}
	\vh^{n+1/2}&=\vh^{n-1/2} -\Delta t \fMmu^{-1} \C \ve^n \\
	\ve^{n+1}&=\ve^{n} -\Delta t \fMeps^{-1}(\Ct\vh^{n+1/2}-\fj_s^{n+1/2})~,
\end{align}
\end{subequations}
which is visualized in Fig.~\ref{Leapfrog}.
The stability of the scheme is connected to the grid dispersion relation, which describes the velocity of a plane wave on the grid as dependent on the direction of the wave vector.
It reads for a particular Cartesian cell $(\Delta x, \Delta y, \Delta z)$
\beq
\left(\frac{\sin \frac{k_x \Delta x}{2}}{\Delta x /2}\right)^2
+\left(\frac{\sin \frac{k_y \Delta y}{2}}{\Delta y /2}\right)^2
+\left(\frac{\sin \frac{k_z \Delta z}{2}}{\Delta z /2}\right)^2
=\mu\eps\left(\frac{\sin\frac{\omega \Delta t}{2}}{\Delta t /2}\right)^2~,
\label{griddispersion}
\eeq
reproducing the continuous dispersion relation
$
k_x^2 + k_y^2 + k_z^2 = \omega^2 \mu \eps
$
in the limit $\Delta x, \Delta y,\Delta z,\Delta t \rightarrow 0$.
In order to fulfil Eq{.}~\eqref{griddispersion} with real valued frequency and wavenumbers, at least
\beq
\label{CFL}
\Delta t \leq \min_{i} \sqrt{\frac{\mu_i \eps_i}{\frac{1}{\Delta x_i^2}+\frac{1}{\Delta y_i^2}+\frac{1}{\Delta z_i^2}}}
\eeq
must hold, where the minimum is taken over all mesh cells. This is also referred to as the Courant--Friedrichs--Lewy (CFL) criterion~\cite{Courant1928} for the time step $\Delta t$. It can be shown, see Ref{.}~\cite{Edelvik2004}, that Eq{.}~\eqref{CFL} is also a sufficient condition for stability on the time step.
Particularly for short bunches, the grid dispersion can lead to unphysical effects, such as a positive wake potential at the head of the bunch or numerical Cherenkov radiation. Therefore dedicated schemes have been developed, which e.g{.} do not have dispersion in the direction of beam propagation~\cite{Zagorodnov2005}.

At low frequencies, the CFL criterion together with the K\"upfm\"uller uncertainty principle pose a strong requirement on the time step.
Since longer wakes have to be computed at low frequencies but the time step is tied on the space step, a large number of time steps need to be computed, which is a massive oversampling of a low-frequency wave.
This is a major drawback of explicit TD methods at low frequencies.

\subsection{Frequency domain}
The MGE can be written in FD as
\begin{subequations}
	\label{MGE_FD}
	\begin{align}
		\label{FDdiscretefaraday}
		\C\vke&=-\text{i}\omega\fkb \\
		\label{FDdiscreteampere}
		\Ct\vkh&=\fkj_s+\fkj+\text{i}\omega\fkd \\
		\label{FDdiscreteecharge}
		\St\fkd&=\qk_s \\
		\label{FDdiscretemcharge}
		\S\fkb&=0~,
	\end{align}
\end{subequations}
which can be combined with the complex material relations to a $3N_p\times3N_p$ SLE
\beq
\left(\Ct\fkMnu\C+\text{i}\omega\fMkap-\omega^2\fMeps \right)\vke=-\text{i}\omega\fkj_s~.
\eeq
Since the FIT is a mimetic discretization, the Helmholtz decomposition can also be applied on the discrete level, preserving the properties discussed in Section~\ref{Helmholtz}.
The monopolar beam source current can be written as ($i_z$ is the $z$-index and $\tilde z_i$ the corresponding coordinate)
\beq
\fkj_{s,z}^{\mathrm{mono}}(i_z)=\int \Jvc_s\cdot \mathrm{d}\vec{A}_{z}=q \text{e}^{-\text{i}\omega \tilde z_i/v}
\label{FITbeamcurrent}
\eeq
and a dipole source is obtained by $\fkj^{\mathrm{dip}}_{s,z}(i_z)=\fkj_{e,z}^{\mathrm{mono}}(x=-d_x/2)-\fkj_{e,z}^{\mathrm{mono}}(x=+d_x/2)$.
For the entry and exit of the beam in the computational domain dedicated boundary conditions are required.
A simplified way to implement these is to use Floquet boundary conditions, i.e{.} to connect the entry and the exit of the beam with the proper phase advance $\exp(\text{i}\omega L/v)$, where $L$ is the length of the computational domain. This has the same effect as if the structure would be repeated infinitely often in a chain.
It is valid for frequencies below the beam pipe cutoff, i.e{.} when the subsequent (repeated) structures do not interact with each other.


The impedances can be evaluated according to Eqs.~\eqref{ZlVol} and \eqref{ZtVol} as
\begin{align}
\Zc_\parallel(\vke(\omega))&=-\frac{1}{q^2}\vke\cdot\fkj^*_\mathrm{mono} \\
\Zc_\perp(\vke(\omega))&=-\frac{v}{\omega(qd_x)^2}\vke\cdot\fkj^*_\mathrm{dip}~,
\end{align}
which can be seen as functionals of the discrete solution of Maxwell's equations in FD.

In FD one cannot use the advantages of diagonal material matrices, and therefore FEM on unstructured meshes can be more advantageous than FIT.
This becomes particularly clear in the discretization of dipole sources for non-ultrarelativistic beams, cf{.} Eq{.}~\eqref{sigma}.
We will briefly discuss a 2D FEM approach which is particularly useful for beam pipes and long kicker magnets.

The finite element method is based on decomposing the computational domain $\Omega$ in finite-sized subdomains $\Omega_e$, i.e{.} the elements.
A function in an appropriate space can be approximated by a finite basis, such that each element is the support of one basis (ansatz) function.
Since such an approximation is (weakly) differentiable only once, a second-order PDE has to be brought in a `weak formulation'.
This is obtained by multiplying with all test functions\footnote{In the standard (Galerkin) approach, the test functions are identical to the ansatz functions.} of an appropriate test-function space, integrating over the whole domain and transferring one (exterior) derivative by means of partial integration. Finally an SLE is obtained, which has number of ansatz functions as columns and number of test functions as rows.

\begin{figure}[t]
    \centering
    \includegraphics[width=0.35\textwidth]{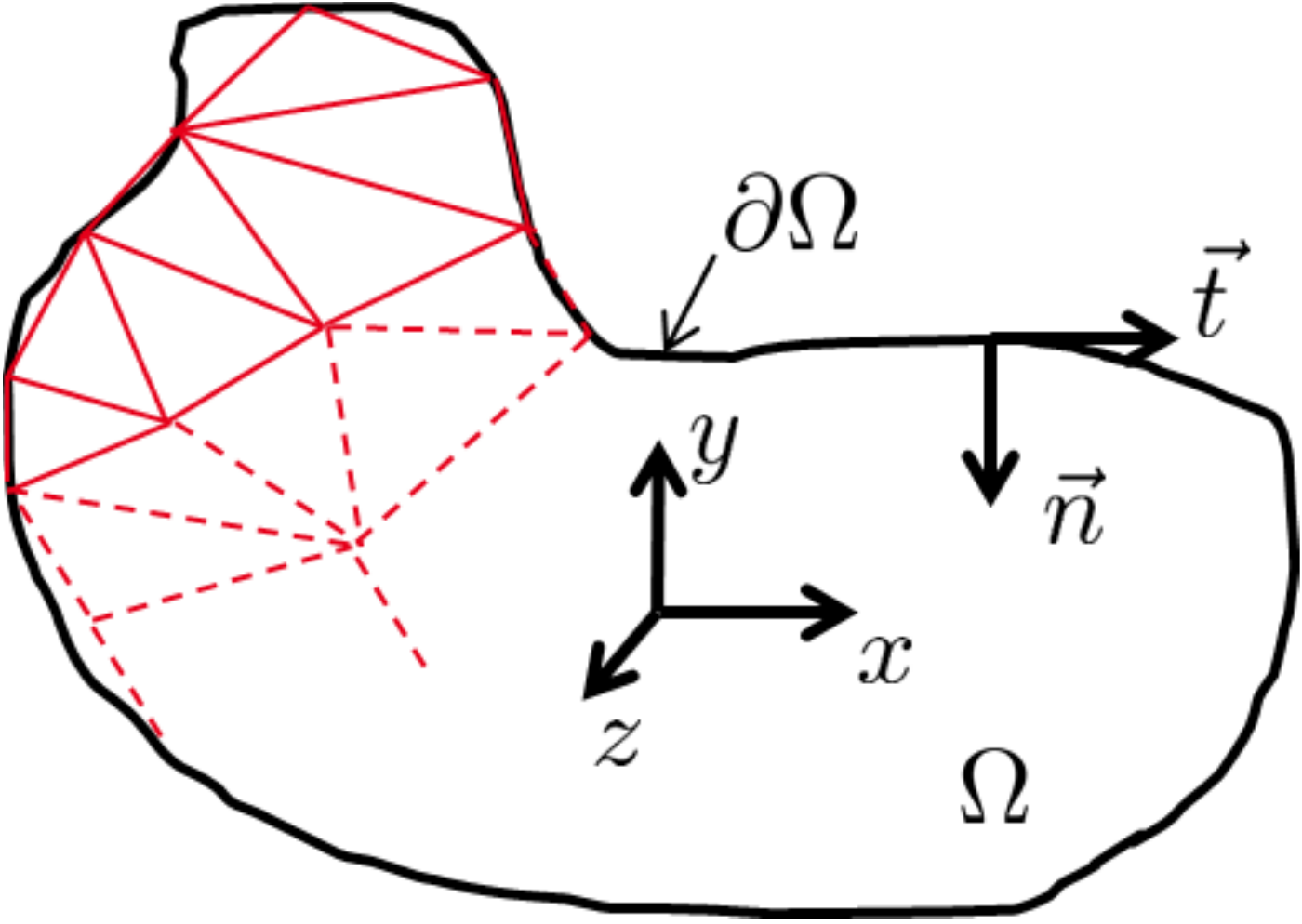}
    \caption{Computational domain for the 2D impedance solver}
    \label{coord}
\end{figure}

We consider again Eq{.}~\eqref{curlcurl}, but $\Evc: \Omega\rightarrow \mathbb{C}^3$ with $\Omega\subset\mathbb{R}^2$ being a simply connected domain as shown in Fig{.}~\ref{coord}.
Each triangle shall be uniformly filled with material $\muc,\epsc$, where $\epsc=\eps_0\eps_r-\text{i}\kappa/\omega$. In order to allow the normal component of the electrical field to jump on a material border while the tangential is continuous, N\'ed\'elec edge elements are employed. These elements (at lowest order) are not suited to model the divergence of a field. Thus a Helmholtz split needs to be performed, i.e{.} Eqs.~\eqref{CurlcurlHelmholzSplit} are discretized.

The discretization of the complex Poisson equation Eq.~\eqref{Poisson} is done by nodal functions\footnote{For simplicity we consider the ansatz functions as purely real, i.e{.} the PDEs need to be split into real and imaginary part, which are coupled in the presence of losses.}
\beq
N_k(\xi,\eta)=a_k\xi+b_k\eta+c_k~,
\eeq
which fulfil
\beq
N_k(\xi_i,\eta_i)=\delta_{i,k}~,
\eeq
where $i$ and $k$ are local vertex indices and $(\xi, \eta)$ are local coordinates.
The lowest-order N\'ed\'elec edge elements of the first kind can be obtained from the nodal elements by (see e.g{.} Ref{.}~\cite{Ingelstroem2006})
\beq
\vec w_i(\xi, \eta)=N_k \nabla\!\!_\perp N_l-N_l\nabla\!\!_\perp N_k~,
\label{Nedelec}
\eeq
and fulfil
\beq
\frac{1}{|l_k|}\int_{l_k} \vec{w}_i\cdot\vec t_k \,\ds=\delta_{i,k}
\eeq
with $\vec t_k$ being the tangential unit vector of edge $l_k$, which is located at the opposite of vertex $k$.
Instead of going further into details about impedance computation with FEM, we refer to \cite{Niedermayer2015b} and references therein.

\begin{figure}[bt]
	\centering
   \includegraphics[width=0.7\textwidth]{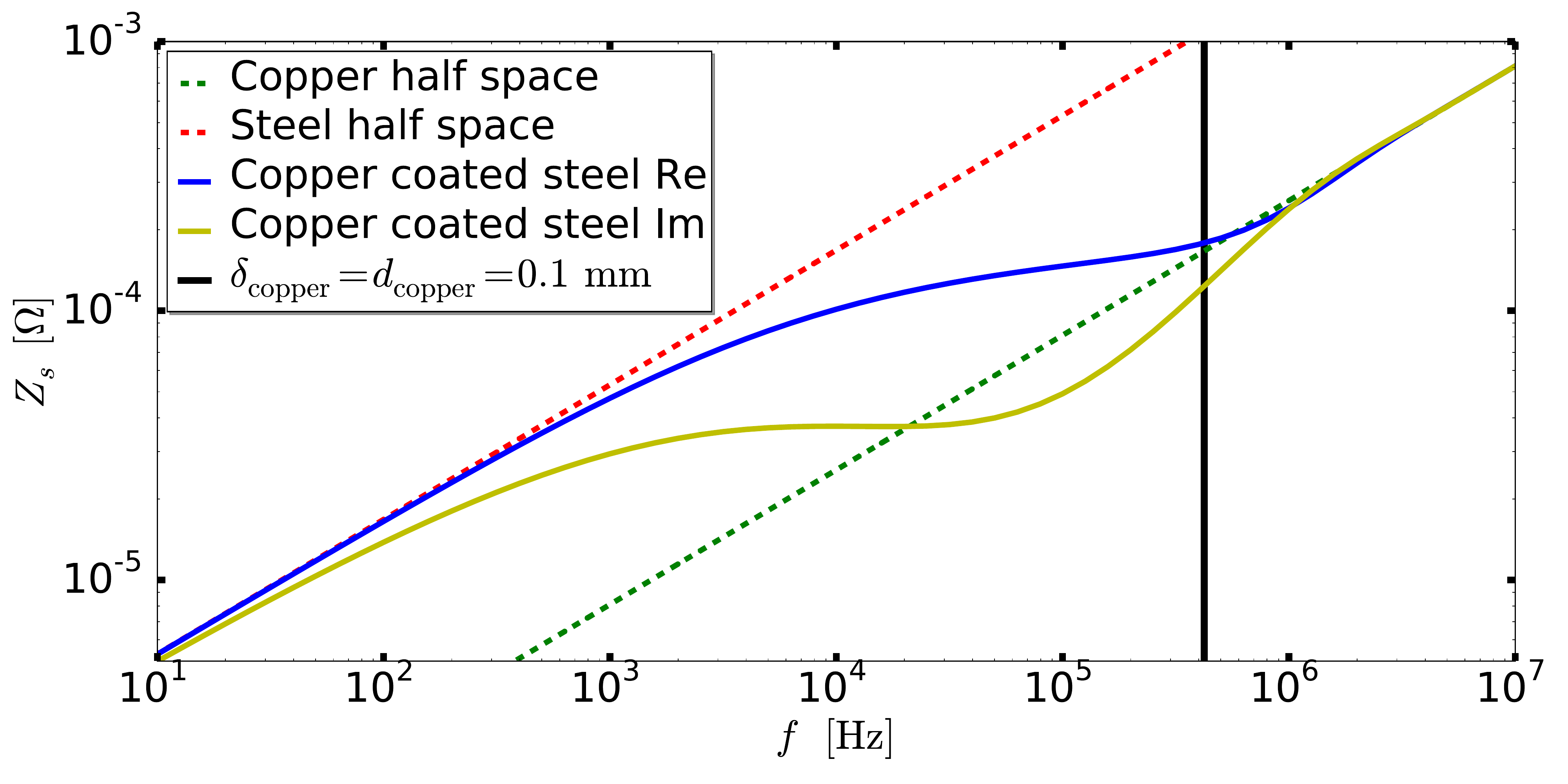}
    		\caption{Surface impedance for a thick steel surface, coated by a thin copper layer}
    		\label{CoatedSurfImp}
\end{figure}

In FD and particularly FEM it is also fairly simple to include a thick conducting wall by means of a surface impedance boundary condition (SIBC), i.e{.}
$\nv\times\nv\times\Evc=\Zc_s\nv\times\Hvc$ with the surface impedance $Z_s=\sqrt{\muc/\epsc}$.
For smooth metal surfaces one finds
\beq
\Zc_s=\frac{1+i}{\kappa\delta_s} \;\;\;\;\;\; \mathrm{with\; the\; skin\; depth} 	 \;\;\;\;\;\; \delta_\mathrm{s}=\sqrt{\frac{2}{\mu\kappa\omega}}~.
\label{ZsLeontovich}
\eeq
This can also be generalized for a metal coating (thickness $d$, permeability $\mu_1$ and conductivity $\kappa_1$) on a metal surface (permeability $\mu_2$ and conductivity $\kappa_2$) as
\beq
\Zc_s=\frac{1+i}{\kappa_1\delta_{s1}}
\frac{M^{(+)} \text{e}^{\text{i}k_{z1}d}+M^{(-)} \text{e}^{-\text{i}k_{z1}d}}
{M^{(+)} \text{e}^{\text{i}k_{z1}d}- M^{(-)}\text{e}^{-\text{i}k_{z1}d}}~,
\label{ZsLeontovichCoating}
\eeq
where
\beq
k_{z 1,2}=\frac{1-\text{i}}{\delta_{s1,2}}~, \;\;\;\; M^{(+)}=1+\sqrt{\frac{\mu_1\kappa_2}{\mu_2\kappa_1}}~, \;\;\; M^{(-)}=1-\sqrt{\frac{\mu_1\kappa_2}{\mu_2\kappa_1}}~.
\eeq
The surface impedance of a copper ($\kappa=70$MS) coated steel ($\kappa=1.4$MS) surface is plotted in Fig{.}~\ref{CoatedSurfImp}.

\subsection{Two examples in two dimensions}
The beam-induced heat load in a ferrite kicker module depends crucially on the ferrite yoke gap.
If there is no gap, the magnetic circuit is closed and the longitudinal impedance is much larger.

In the presence of a gap, the longitudinal impedance can be further decreased by increasing the path length for the magnetic field lines outside the ferrite.
Achieving this by increasing the gap thickness can be disadvantageous for the kick-field quality.
The gap can also be filled highly conductive material, i.e{.} copper, which influences the kick field only weakly.

\begin{figure}
	\centering
	\includegraphics[width=\textwidth]{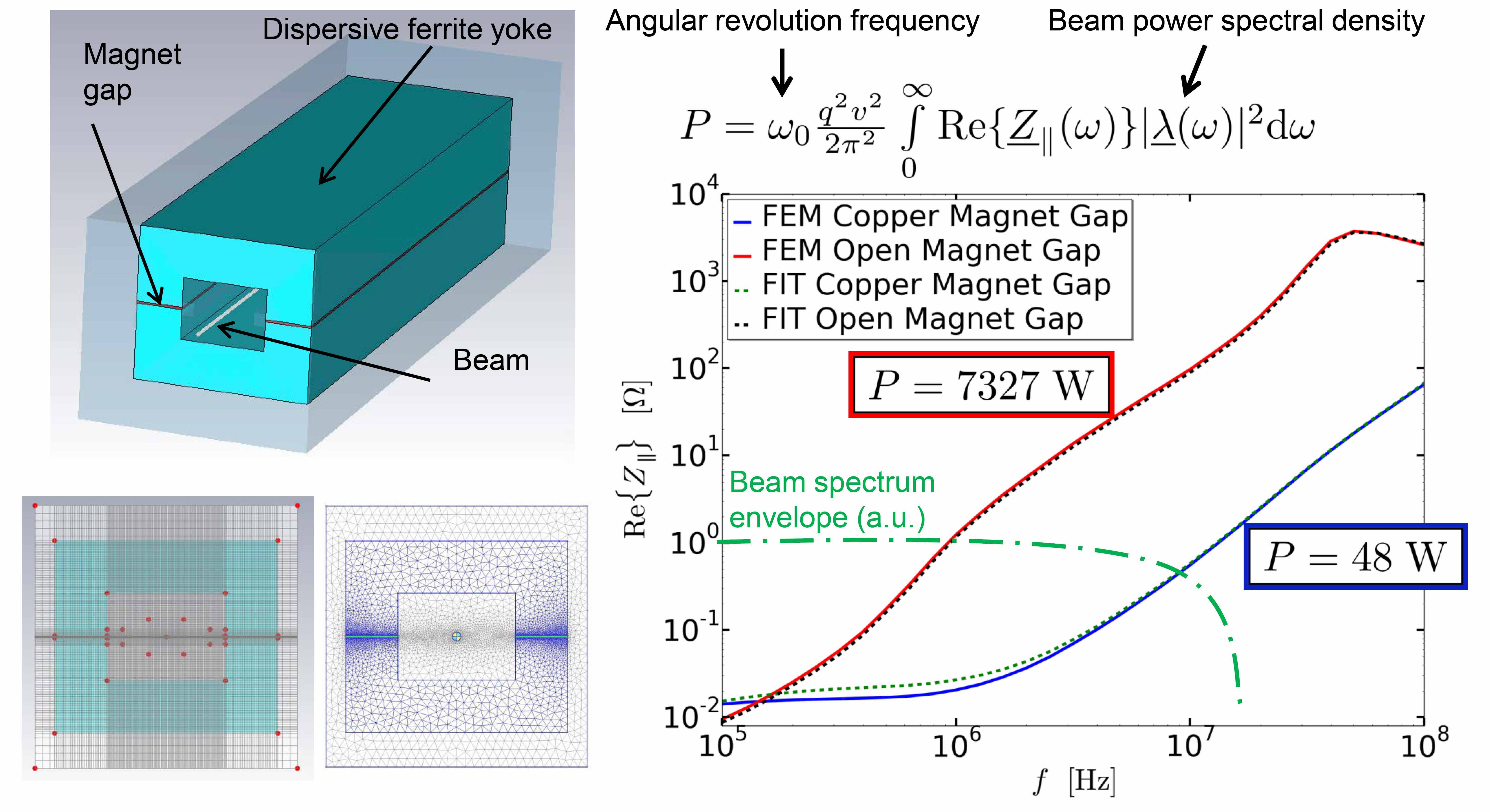}
  \caption{Outline of a GSI SIS-100 transfer kicker. The plot shows the longitudinal impedance computed with FIT on Cartesian and FEM on triangular mesh, for both open and copper filled magnet gaps. }
			\label{2Dkicker}
\end{figure}

Figure~\ref{2Dkicker} shows the longitudinal impedance of a GSI SIS-100 transfer kicker module (design outline) from 2D FIT and FEM simulations, as a comparison between a vacuum and a copper-filled magnet gap.
The difference between the two is significant, i.e{.} two orders of magnitude. The heat power values are for the shortest (3.7~m) single Gaussian bunch with $N=2\times10^{13}$ protons.
Note that these numbers are for CW operation, in practice they have to be scaled with the duty factor which is in the range of 0.5 at most.

\begin{figure}
\centering
   \includegraphics[width=1.0\textwidth]{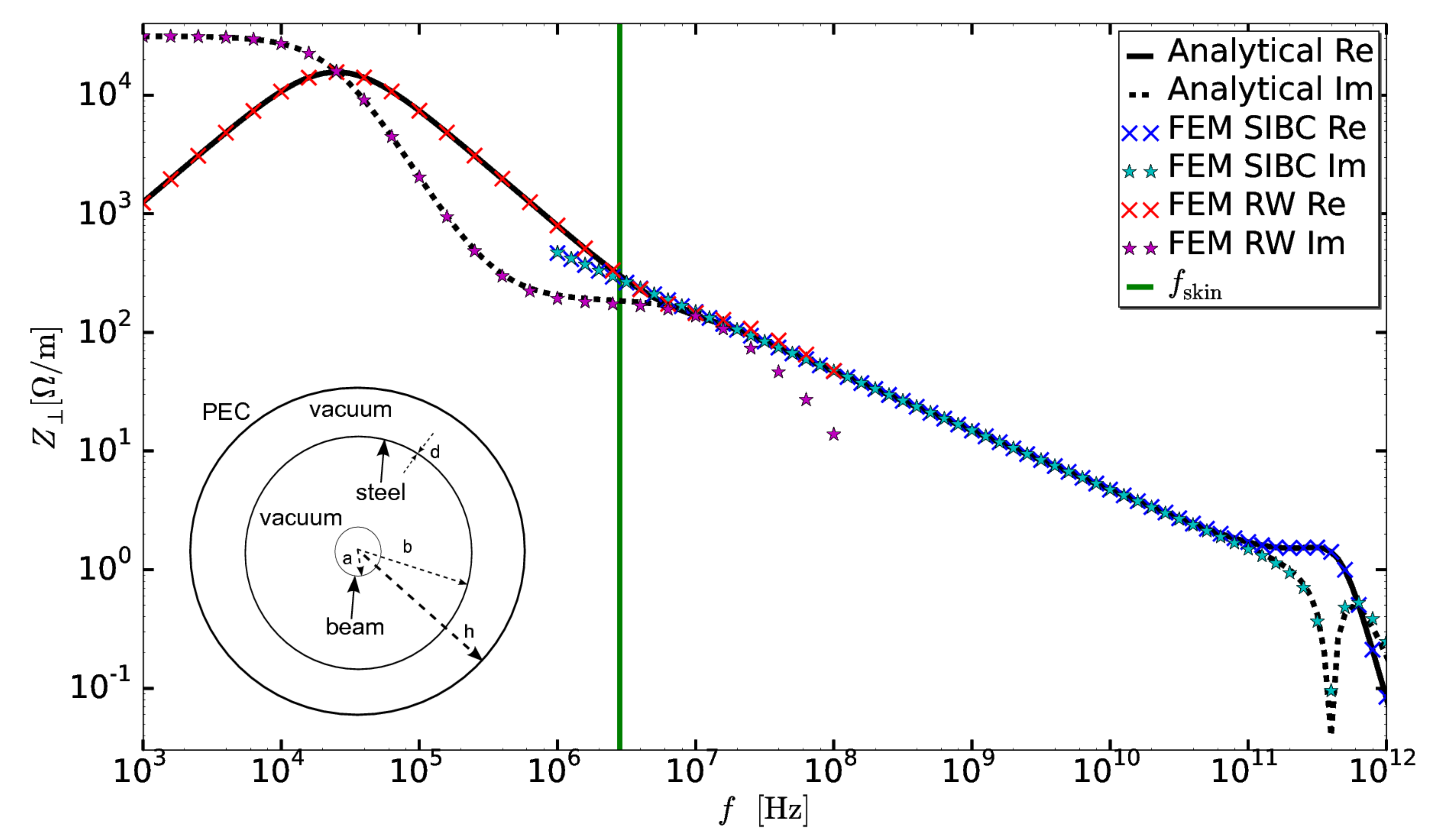}
    		\caption[Transverse impedance of a thin beam pipe.]{Transverse impedance of the thin beam pipe with radius $b=4$\,cm, thickness $d=0.3$\,mm, outer boundary radius $h=1$\,m, length $l=1$\,m, conductivity $\kappa=1$\,MS/m and $\beta=0.999999$.
				The analytical curve originates from Rewall~\cite{Mounet2009}. At low frequencies, the resistive wall is resolved by the mesh (RW-marks, red and magenta), while at high frequency the surface impedance boundary condition was applied (SIBC-marks, blue and cyan).}
				\label{thinpipeZtFEM}
\end{figure}


The second example is the transverse impedance of a thin-layered round beam pipe, see Fig{.}~\ref{thinpipeZtFEM}.
At low frequency the wall has to be meshed since field transmission plays a significant role, especially for frequencies below the maximum.
For frequencies above the skin frequency $f_s$ the skin depth becomes very small and cannot be properly meshed. By means of the SIBC, resolving the skin depth can be avoided and high frequencies can be reached.

\section{Impedance bench measurements}
The transmission-line measurement technique is based on replacing the beam by a wire and measuring the attenuation of a TEM wave.
It was introduced in 1974 by Sands and Rees~\cite{Sands1974} in order to determine the beam energy-loss factors in the TD using a broadband pulse with a shape similar to the particle bunch.
Nowadays, modern vector network analysers (VNA) allow sweeping a narrow-banded sinusoidal signal, to obtain the impedance directly for a particular frequency range.
Especially when particular beam instability sidebands are under investigation, the FD method is advantageous.

The motivation of the wire measurement comes from a the fields of a point charge moving at velocity $\beta c$ which read in the rest frame of the charge as
\beq
\Ev'(\rho',z',t')=\frac{q}{4\pi\varepsilon}\left(\frac{\rho'}{\sqrt{\rho'^2+z'^2}^3}\erho+\frac{z'}{\sqrt{\rho'^2+z'^2}^3}\ez\right)~.
\eeq
Lorentz-boosting to the laboratory frame one obtains
\beq
\Ev(\rho,z,t)=\frac{q}{4\pi\varepsilon}\left(\frac{\gamma\rho}{\sqrt{\rho^2+(\beta\gamma c t)^2}^3}\erho+\frac{-\beta\gamma c t}{\sqrt{\rho^2+(\beta\gamma c t)^2}^3}\ez\right)~,
\;\;\; H_\phi=\frac{\beta}{Z_0}E_\rho
\eeq
and subsequent Fourier transform results in
\begin{subequations}
\begin{align}
\label{EzSource}
\Ec_z(\rho,z,\omega)&=\mathrm{i} q \frac{\mu_0}{2\pi}\frac{\omega}{\beta^2 \gamma^2} \rm{K}_0 \left( \frac{\omega}{\beta\gamma c} \rho \right) \xrightarrow{\gamma \rightarrow \infty}0\\
\label{ErhoSource}
\frac{1}{\beta}Z_0\Hc_\phi=\Ec_\rho(\rho,z,\omega)&=q \frac{\mu_0}{2\pi}\frac{\omega}{\beta^2 \gamma} \rm{K}_1 \left( \frac{\omega}{\beta\gamma c} \rho \right) \xrightarrow{\gamma \rightarrow \infty} q \frac{Z_0}{2\pi\rho}
\end{align}
\end{subequations}
which is a TEM mode in the ultrarelativistic limit.
From Eq{.}~\eqref{ErhoSource}, we can see that the wire technique corresponds only to an ultrarelativistic beam, since the wave impedance for a beam is
\beq
Z_\mathrm{wave}=\frac{\Ec_\rho}{\Hc_\phi}=\frac{Z_0}{\beta},
\eeq
but for a TEM mode one has always $Z_\mathrm{wave}=Z_0$. Moreover, the wire needs to be thin, in order to have most of the field unperturbed close to the wire. In the wide sense, this corresponds to the rigid-beam approximation.



\subsection{Basics of RF vector network analysis and impedance matching}
\begin{figure}[t!]
	\centering
		\includegraphics[width=0.85\textwidth]{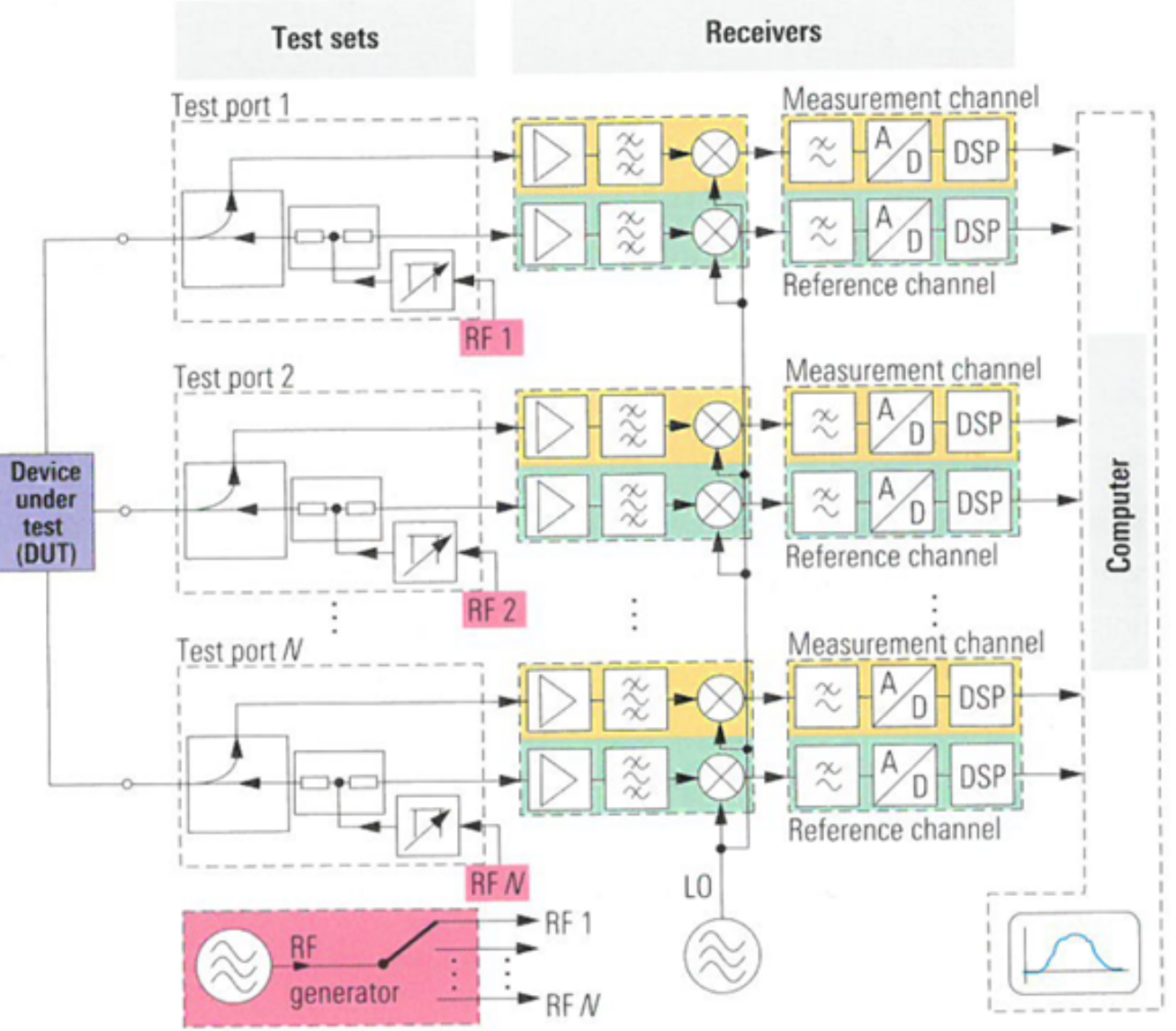}
    	\caption{Block schematic of a modern VNA. Picture adapted from Ref{.}~\cite{Hiebel}
			}
    		\label{VNABlock}
\end{figure}

\begin{figure}[h!]
	\centering
   \includegraphics[width=0.8\textwidth]{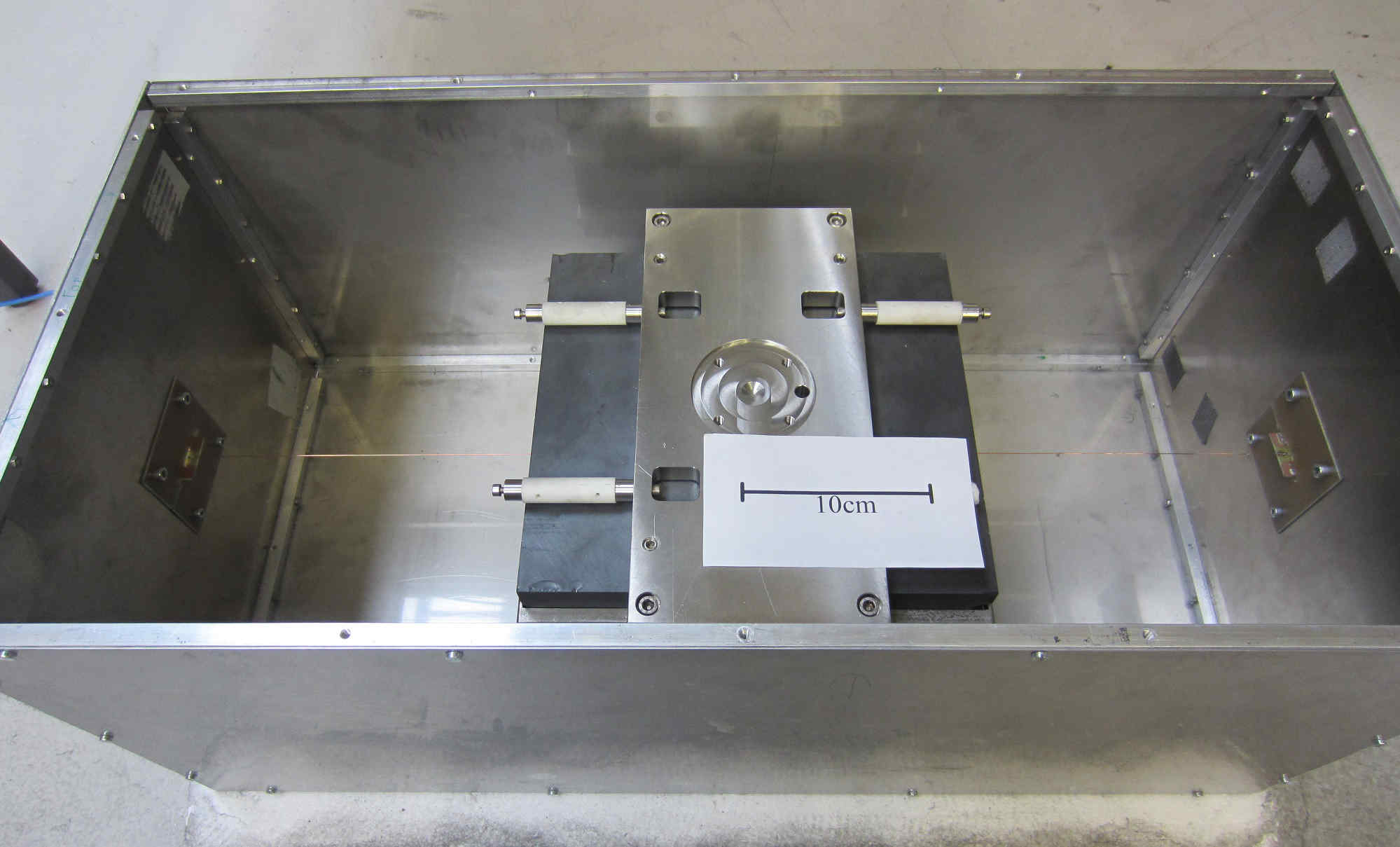}
	\includegraphics[width=0.8\textwidth]{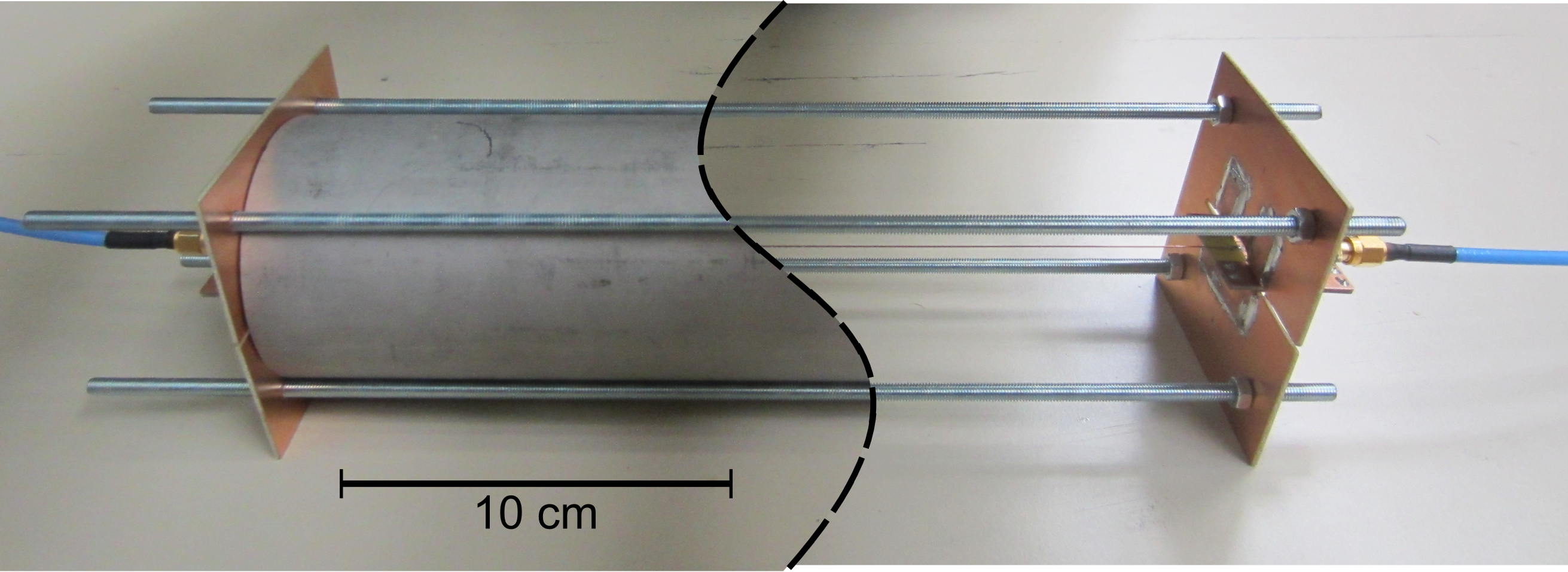}
    	\caption{Large ($Z_c=433\pm 18\, \Omega$) and small ($Z_c=299\pm 12\, \Omega$) measurement box
			}
    		\label{setup_meas1}
\end{figure}

\begin{figure}[h!]
	\centering
	\subfloat[Matching with a simple voltage divider.]{
		\label{ResistiveMatching}
      \includegraphics[width=0.53\textwidth]{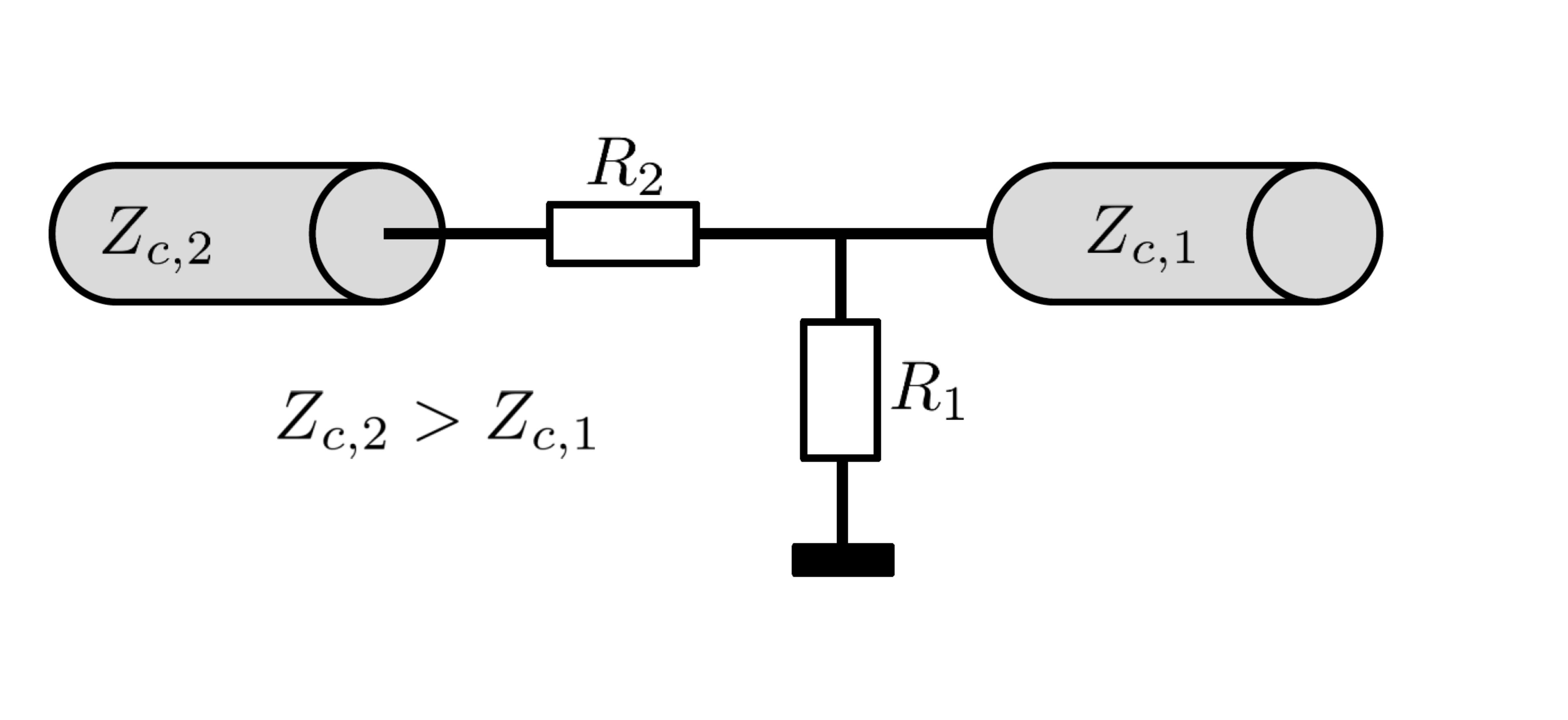}
	}
	\hfill
	\subfloat[Matching with resistors and a 10~dB attenuator~\cite{Niedermayer2015}.]{
	\label{attenuator}
	\includegraphics[width=0.3\textwidth]{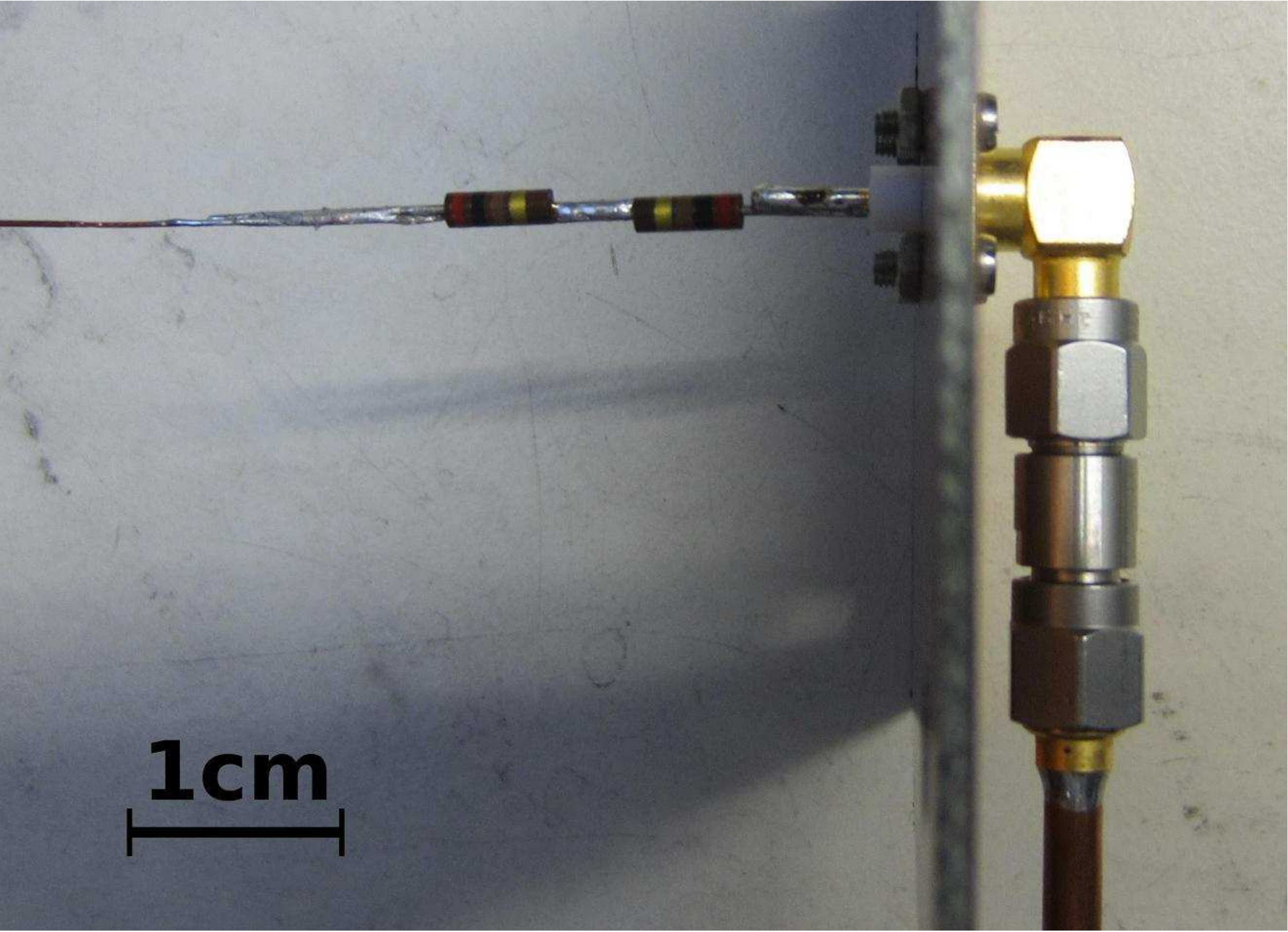}
				}
				\caption{Different types of resistive matching}
    		\label{matching}
\end{figure}

In Radio Frequency (RF) systems, the integral of the electric field strength depends on the taken path, thus voltages are not uniquely defined. To account for that, power flow parameters
\beq
a_i:=\frac{1}{2\sqrt{Z_c}}(\Uc_i+Z_c \Ic_i)~, \;\;\;b_i:=\frac{1}{2\sqrt{Z_c}}(\Uc_i-Z_c \Ic_i)
\eeq
in units of $\sqrt{\mathrm{W}}$ are used rather than voltages and currents. The out-flowing power of a linear, time-invariant device with $i=1,\ldots, N$ ports is determined from the inflowing power by $\vec b= \mathbf{S} \vec a$,
where the scattering parameters $S_{ij}$ are defined by
\beq
S_{ij} =\left. \frac{b_i}{a_j}\right|_{a_k=0\;\; \forall k\neq j}~.
\eeq
For simplicity, we assume the characteristic impedance for all ports to be equal and consider only the TEM mode.
The scattering matrix can be determined by numerical simulation or by measurement with a vector network analyser (VNA), see Fig{.}~\ref{VNABlock} for a block schematic of a VNA.
Prior to measurement, the VNA needs to be calibrated at a particular reference plane close to the device under test (DUT), in order to exclude effects of the cables.
Phase-stable cables are crucial for measurements at higher frequencies, but such precision measurement cables are costly.
A good alternative are semi-rigid cables, which maintain their shape and therefore have only little phase drift.
In order to connect the $50\,\Omega$ lines of the VNA with the wire for beam impedance measurement, the characteristic impedance~\cite{Pozar}
\beq
\label{Zchar}
Z_c=\frac{Z_0}{2\pi}\ln\frac{b}{a}~,
\eeq
where $b$ and $a$ are the radii of the outer and inner conductor of a coaxial cable, needs to be matched. A mismatched transition would lead to a reflection~\cite{Pozar}
\beq
r=\frac{Z_{c,1}-Z_{c,2}}{Z_{c,1}+Z_{c,2}}
\eeq
which is for the transition of a thin wire in a measurement box (as depicted in Fig{.}~\ref{setup_meas1}) to a conventional $50\,\Omega$ line in the range of 80\%.
Thus, a matching network is required, since otherwise multiple reflections would be in the same range of amplitude as the primary signal.
The simplest way to construct a matching network is to use RF-resistors (carbon composite) in a way that each side sees its own characteristic impedance, e.g{.} as depicted in Fig{.}~\subref*{ResistiveMatching}.
\\
Here, the two resistors $R_1$ and $R_2$ have to fulfil the two matching conditions
\begin{subequations}
\label{matchingconditions}
\begin{align}
R_1\parallel(R_2+Z_{c,2})&=Z_{c,1}\\
R_2+R_1\parallel Z_{c,1}&=Z_{c,2}~,
\end{align}
\end{subequations}
where $x\parallel y=xy/(x+y)$ is the abbreviation for parallel circuits.
Unfortunately, practical resistors have an inductance, which makes it impossible to fulfil Eqs.~\eqref{matchingconditions} at higher frequencies in a broadband manner.
This can be partly overcome by taking an attenuator instead of $R_1$, see Fig{.}~\subref*{attenuator}. Commercially available attenuators are broadband matched to $50\,\Omega$ and do not suffer from the inductance problem.

\begin{figure}[t]
	\centering
	\subfloat[REF measurement in order to de-embed the DUT.]{
   \includegraphics[width=0.5\textwidth]{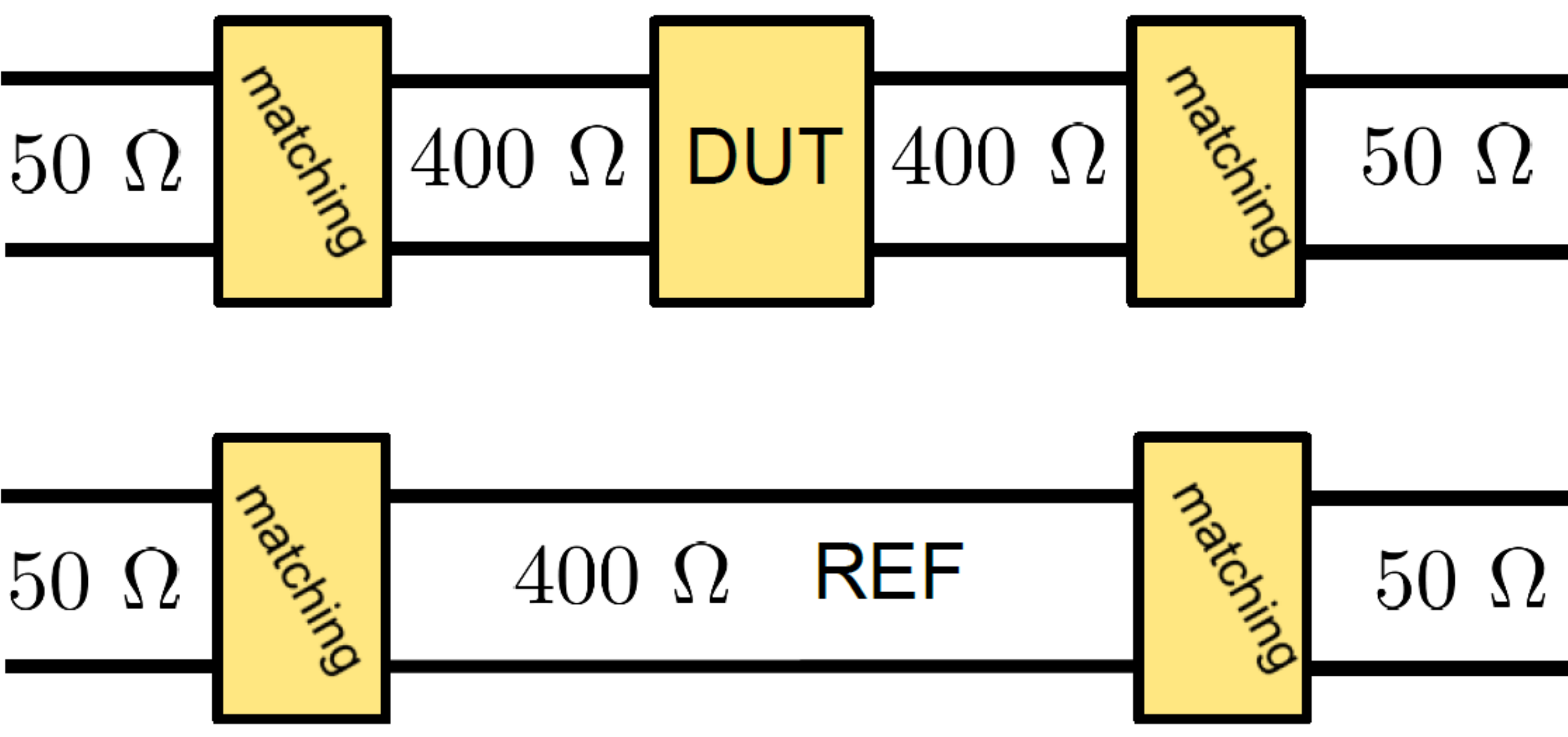}
		}
		\hspace{1cm}
	\subfloat[Quantities of interest for the de-embedded DUT.]{
		\label{DUTde-embed}
	   \includegraphics[width=0.3\textwidth]{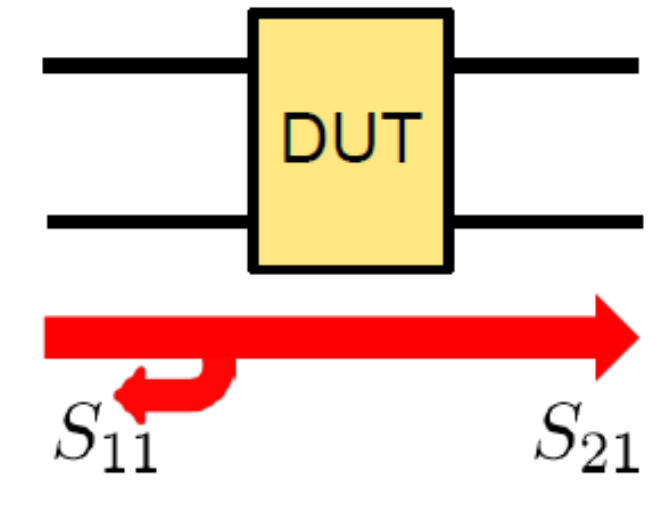}
		}
				\caption{De-embedding by subsequent DUT and REF measurements}
				\label{deembedding}
\end{figure}

The frequency-dependent attenuation and phase shift of the matching network is calibrated out by a reference measurement, such that for an assumed perfect matching the de-embedded transmission is
$S_{21}^{\rm de-embed}(\omega)=S_{21}^\mathrm{DUT}(\omega)/S_{21}^\mathrm{REF}(\omega)$.
Here, only the reflection of the matching network needs to be close to zero, but it is allowed to be lossy, within the dynamic range of the VNA.
The way how $S_{21}^\mathrm{DUT}$ and $S_{21}^\mathrm{REF}$ are measured is illustrated in Fig{.}~\ref{deembedding}.
Obviously, it is advantageous to have a setup as small as possible, in order to shift the eigenmodes (resonances) of the box to frequencies as high as possible.


\subsection{Wire method}
There is a crucial difference between the beam and the wire setup: the TEM wave experiences an
attenuation, which is not negligible and actually the quantity to be measured by the $S_{21}$-parameter. Thus,
lumped (short) and distributed (long) impedances require different interpretations of the
measured $S_{21}$-parameters. Mathematically, lumped and distributed impedances can be identified by their distribution along the $z$-axis, i.e{.}
\begin{subequations}
\label{impedancedistribution}
\begin{align}
\frac{\partial \Zc_\parallel^\mathrm{lumped} (\omega,z)}{\partial z}&=\Zc_\parallel^\mathrm{total}(\omega) \delta(z-z_0)\\
\frac{\partial \Zc_\parallel^\mathrm{dist} (\omega,z)}{\partial z}&=\frac{\Zc_\parallel^\mathrm{total}(\omega)}{l}~.
\end{align}
\end{subequations}
In a real accelerator components, there is always a mixture of both.
The impedance discontinuity (geometric impedance) at the beginning of the DUT is always lumped, while the body of the DUT (resistive wall) is often almost equally distributed.

\begin{figure}[h!]
\makebox[.5\textwidth]{
	\subfloat[Lumped impedance]{
		\label{Lumped}
		\centering
      \includegraphics[width=0.35\textwidth]{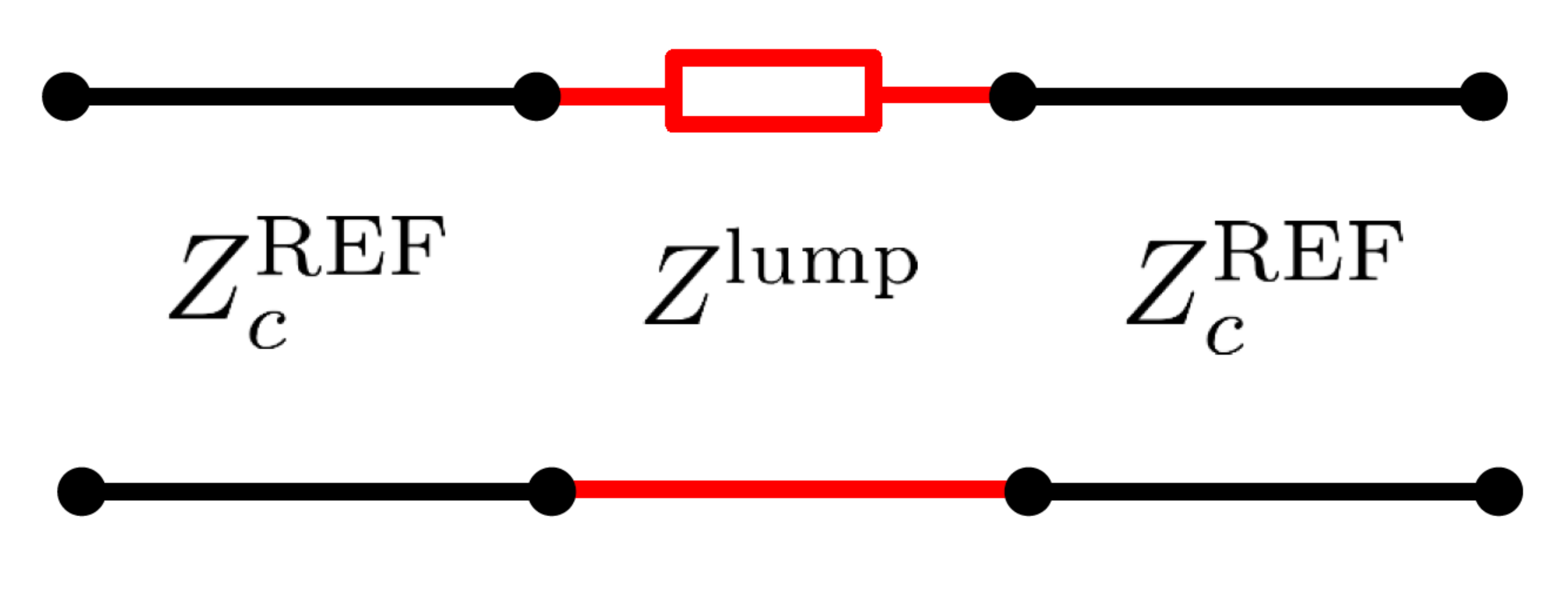}
	}
	}
	\makebox[.5\textwidth]{
	\subfloat[{Distributed~impedance}]{
	\label{Dist}
		\centering
	\includegraphics[width=0.35\textwidth]{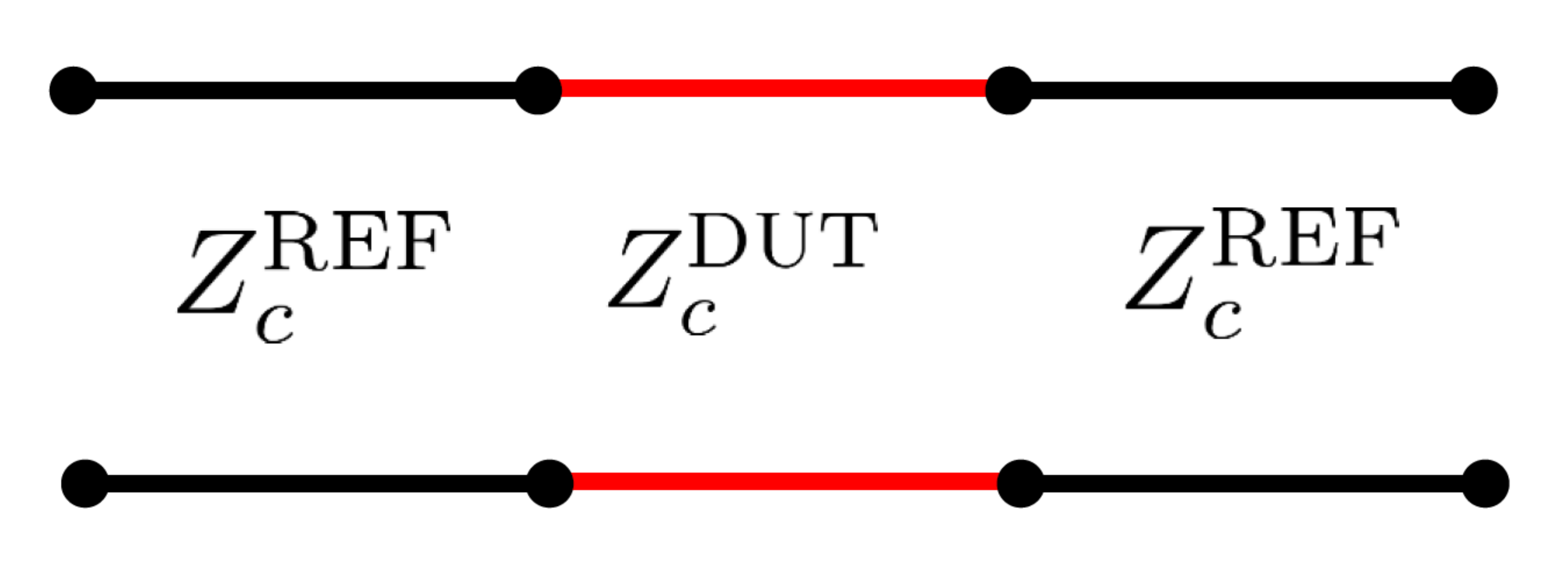}
				}
				}
				\caption{Different modelling approaches for the de-embedded accelerator component}
    		\label{LumpedVSDist}
\end{figure}

The modelling of lumped impedances is just a localized impedance element in the longitudinal direction, while distributed impedances can be represented by a TEM line with an impedance element $Z_\parallel/l$ equally distributed to each infinitely short transmission-line element, see Fig{.}~\ref{LumpedVSDist}. We call a device a lumped (distributed) impedance if $\Zc^\mathrm{lumped}\gg \Zc^\mathrm{dist}$ ($\Zc^\mathrm{lumped}\ll \Zc^\mathrm{dist}$).

The scattering matrices for the de-embedded DUT are given by Ref{.}~\cite{Pozar, Vaccaro1994}
\begin{align}
	\mathbf{S}^\mathrm{lump}
	=\frac{1}{2Z_c^\mathrm{REF}+Z^\mathrm{lump}}	\left(
	\begin{matrix}
	 Z^\mathrm{lump} & 2Z_c^\mathrm{REF}\\
	 2Z_c^\mathrm{REF} & Z^\mathrm{lump}
	\end{matrix}
	\right)
 \label{S_lump}
\end{align}
\beq
	\mathbf{S}^\mathrm{dist}
	=\frac{	\left(
	\begin{matrix}
	(Z_c^\mathrm{DUT^2}-Z_{c}^\mathrm{REF^2})\sin(k_\mathrm{z}^\mathrm{DUT}l) & -2iZ_c^\mathrm{DUT}Z_{c}^\mathrm{REF}\\
	-2iZ_c^\mathrm{DUT}Z_{c}^\mathrm{REF} & (Z_c^\mathrm{DUT^2}-Z_\mathrm{c}^\mathrm{REF^2})\sin(k_\mathrm{z}^\mathrm{DUT}l)
	\end{matrix}
	\right)}
	 {(Z_c^\mathrm{DUT^2}+Z_{c}^\mathrm{REF^2})\sin(k_\mathrm{z}^\mathrm{DUT}l)-2iZ_c^\mathrm{DUT}Z_{c}^\mathrm{REF}\cos(k_\mathrm{z}^\mathrm{DUT}l)}
 \label{S_dist}
\eeq
for the lumped and distributed impedance, respectively.
How to derive the DUT impedance from the measured $S$-parameters will be discussed for lumped and distributed impedances in the following subsections.


\subsubsection{Distributed-impedance measurement}
\begin{figure}[h]
	\centering
	\subfloat[REF]{
   \includegraphics[width=0.45\textwidth]{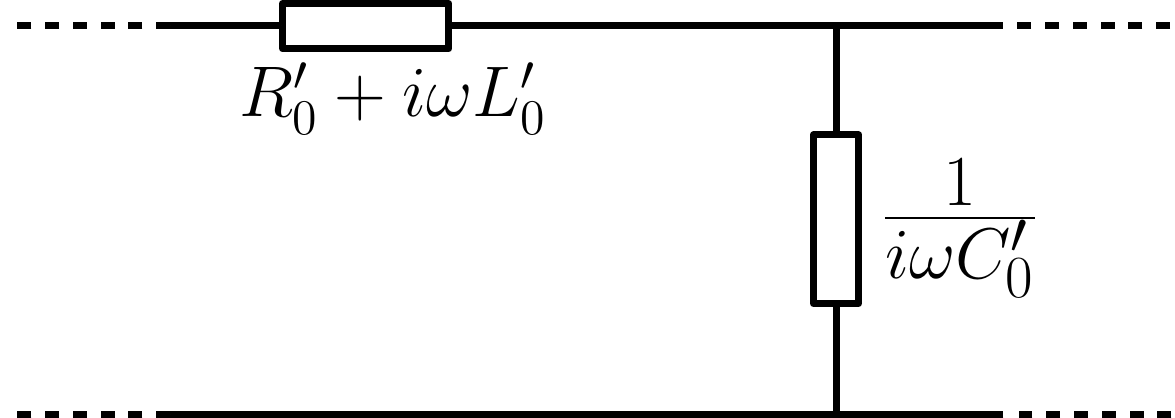}
	}
	\hspace{1cm}
		\subfloat[DUT]{
   \includegraphics[width=0.45\textwidth]{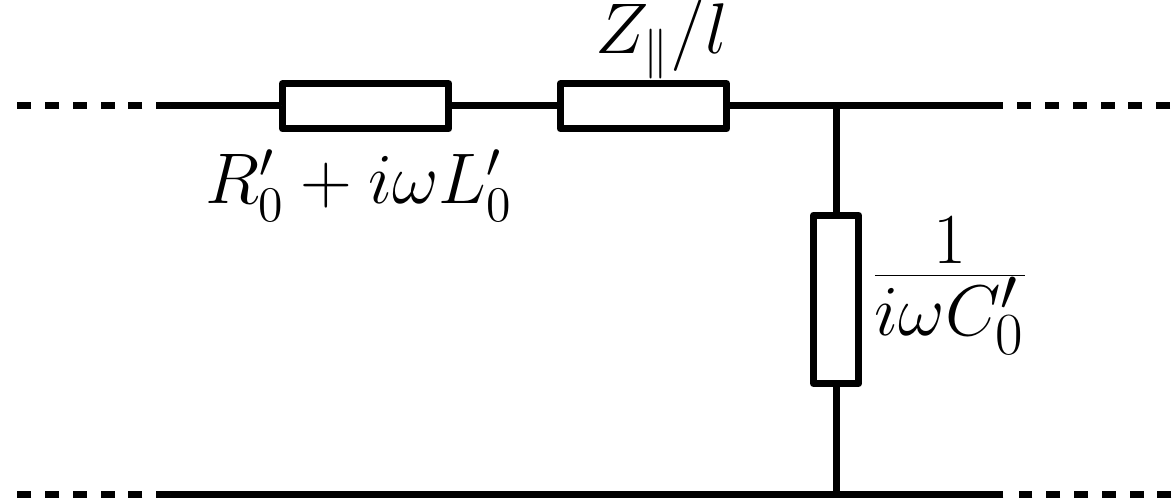}
	}
    		\caption{Transmission-line replacement circuit for distributed impedance}
    		\label{equivalentcircuit_all}
\end{figure}

In transmission-line theory, a ladder-replacement-circuit model as shown in Fig{.}~\ref{equivalentcircuit_all} can be derived. Here, $L_0',C_0'$, and $R_0'$ are inductance, capacitance, and resistance per length, respectively.
The distributed beam coupling impedance can be seen as an additional longitudinal element $Z_\parallel/l$.


From the transmission-line parameters the propagation constants and characteristic impedances can be calculated as (see Ref{.}~\cite{Pozar})
\begin{subequations}
\begin{align}
	k_\mathrm{z}^\mathrm{DUT}&=\omega \sqrt{C'_0 L'_0} \sqrt{ 1-\text{i}\frac{R'_0+\Zc_\parallel/l}{\omega L'_0}}  \\
	k_\mathrm{z}^\mathrm{REF}&=\omega \sqrt{C'_0 L'_0} \sqrt{ 1-\text{i}\frac{R'_0}{\omega L'_0}}  \\
	\Zc_{c}^\mathrm{DUT} &=\sqrt{\frac{R'_0 +\text{i}\omega L'_0+\Zc_\parallel/l}{\text{i}\omega C'_0}}    \\
	\Zc_{c}^\mathrm{REF} &=\sqrt{\frac{R'_0 +\text{i}\omega L'_0}{\text{i}\omega C'_0}}\approx\sqrt{\frac{L'_0}{C'_0}} =:Z_c.
\end{align}
\end{subequations}
This system can be solved for $Z_\parallel$ as
\beq
	\Zc_\parallel^\mathrm{dist}= i Z_c^\mathrm{REF}l\cdot (k_\mathrm{z}^\mathrm{DUT}-k_\mathrm{z}^\mathrm{REF})\cdot \left( 1+ \frac{k_\mathrm{z}^\mathrm{DUT}}{k_\mathrm{z}^\mathrm{REF}}\right) .
	\label{Zlongfromk}
\eeq
Since the DUT setup is a combination of three transmission lines (cf{.}~Fig{.}~\ref{LumpedVSDist}), obtaining the propagation constants can be involved, when a reflection takes place at the DUT.
When this reflection is small, i.e{.} $Z_c^\mathrm{DUT}\simeq Z_c^\mathrm{REF}$~, Eq{.}~\eqref{S_dist} simplifies to
\beq
	S_{21}=S_{12}=\text{e}^{-\text{i}k_\mathrm{z} l}\;, \;\;  S_{11}=S_{22}=0~,
	\label{equ_S_assumption}
\eeq
which can be easily inverted.
Otherwise a reflection-corrected $S_{21}$-parameter can be introduced, which is by definition
\beq
\label{S21C}
S_{21}^\mathrm{C}:=\text{e}^{-\text{i}k_\mathrm{z} l}~.
\eeq
The new $S_{21}^\mathrm{C}$-parameter can be obtained by solving Eq{.}~\eqref{S_dist} for $k_\mathrm{z}$, which can be achieved through replacing sine and cosine by exponentials.
The quadratic equation thereby derived for $S_{21}^\mathrm{C}$ is called the Wang--Zhang formula~\cite{Wang2000} and is
\begin{align}
 \label{WangZhang}
	(S_{21}^\mathrm{C})^2+\frac{S_{11}^2-S_{21}^2-1}{S_{21}}S_{21}^\mathrm{C} +1 =0  ~,
\end{align}
where only one of the two solutions, which fulfils $|S_{21}^\mathrm{C}|<1$, is physical.
Solving Eq{.}~\eqref{WangZhang} requires the knowledge of the $S_{11}$-parameter, which is in practice difficult to measure because of multiple reflections between the DUT and the matching section (cf{.} Fig{.}~\ref{deembedding}).
Nonetheless, $S_{11}$ can be determined easily in simulations with waveguide ports.

The wavenumber $k_\mathrm{z}$ is found from the complex logarithm of either the original (see Eq{.}~\eqref{equ_S_assumption}) or the corrected (see Eq{.}~\eqref{S21C}) $S_{21}$-parameter.
It can be inserted into Eq{.}~\eqref{Zlongfromk} to obtain (see Ref{.}~\cite{Vaccaro1994})
\beq
	\Zc_\parallel^\mathrm{dist}=Z_c \cdot \ln \left(\frac{S_{21}^\mathrm{REF}}{S_{21}^\mathrm{DUT}}\right)\cdot \left[1+\frac{\ln (S_{21}^\mathrm{DUT})}{\ln (S_{21}^\mathrm{REF})}\right]~,
	\label{improved-log-formula}
\eeq
which is called `improved-log-formula'\footnote{Historically, first the lumped-element formulas~\cite{Sands1974,Hahn1978}, then the simplified transmission-line formula `Log-formula'~\cite{Walling1989}, and finally the full transmission-line formula `Improved Log-formula'~\cite{Vaccaro1994} were derived.} in the literature.
This formula is exact for ideally distributed impedances, but it does not apply to lumped impedances, since the replacement circuit in Fig{.}~\ref{equivalentcircuit_all} requires many such transmission-line elements in succession. The dependence on the electrical length of the reference $\Theta_\mathrm{z}^\mathrm{REF}=k_\mathrm{z}^\mathrm{REF}l=\omega l/c$ can be pointed out explicitly by rewriting
Eq{.}~\eqref{improved-log-formula} as (see Ref{.}~\cite{Jensen2000})
\beq
\Zc_\parallel^\mathrm{dist}=Z_c \cdot \ln \left(\frac{S_{21}^\mathrm{REF}}{S_{21}^\mathrm{DUT}}\right)\cdot \left[2+\frac{i}{\Theta_\mathrm{z}^\mathrm{REF}}\ln\left(\frac{S_{21}^\mathrm{DUT}}{S_{21}^\mathrm{REF}}\right)\right]~. \label{improved-log-formula2}
\eeq
This formula contains only the logarithm of the ratio, i.e{.} the difference term in Eq{.}~\eqref{Zlongfromk}.
Besides the implicit dependence of $k_z^\mathrm{DUT}$ on $l$, the  $\ln(S_{21}^\mathrm{DUT}/S_{21}^\mathrm{REF})$-term is linear in $l$.
Thus, for distributed impedances, the square bracket in Eq{.}~\eqref{improved-log-formula2} does not depend on the length explicitly.

When inserting the lumped impedance $S$-parameters (see Eq{.}~\eqref{S_lump}) into the improved-log-formula (see Eq{.}~\eqref{improved-log-formula} or \eqref{improved-log-formula2}) one observes that the second term in the bracket is not independent of the length anymore, i.e{.} $\Theta_z$ does not cancel. This shows explicitly the inapplicability of the improved-log-formula to lumped impedances since the length is not defined for lumped impedances.
In other words, if a differentially short transmission-line element is assigned a finite impedance value (lumped impedance) and this is integrated over a finite length, then the result must diverge.
The convergence of $\Zc_\parallel^\mathrm{dist}$ to $\Zc_\parallel$ in the limit of decreasing wire radius is discussed in Ref{.}~\cite{Niedermayer2015}.


\subsubsection{Lumped impedance measurement}
The determination of lumped impedances is significantly simpler than the one for distributed impedances, since the reflection does not influence the transmission measurement result.
In fact, the reflection can even be used as an alternative method to determine a lumped impedance. However, Hahn and Pedersen argued, in Ref{.}~\cite{Hahn1978}, that the reflection method is inferior to the transmission method.
From solving Eq{.}~\eqref{S_lump} for $\Zc^\mathrm{lump}$ one obtains the so-called Hahn--Pedersen lumped impedance formula \cite{Hahn1978},
\beq
   \Zc_{\parallel, \mathrm{HP}}^\mathrm{lump}=2 Z_c\frac{S_{21}^\mathrm{REF}-S_{21}^\mathrm{DUT}}{S_{21}^\mathrm{DUT}}~.
	\label{HPformula}
\eeq
In modern VNAs this impedance measurement formula is already built-in, i.e{.} the impedance can be directly displayed for the simplified case $S_{21}^\mathrm{REF}=1$.
Equation~\eqref{HPformula} is an improvement of the original Sands--Rees pulse-energy-loss formula~\cite{Sands1974} (see also Ref{.}~\cite{Karantzoulis1991})
\beq
   \Zc_{\parallel, \mathrm{SR}}^\mathrm{lump}=2 Z_c\frac{S_{21}^\mathrm{REF}-S_{21}^\mathrm{DUT}}{S_{21}^\mathrm{REF}}~.
	\label{SRformula}
\eeq
Note that there is no theoretical limit on the impedance magnitude for the determination of purely lumped impedances.
A proof, that the measured lumped impedance converges to the beam impedance for decreasing wire radius, is outlined in \cite{Argan1999}.

\subsubsection{Mixed-impedance measurement}
Both the lumped (Hahn--Pedersen) and the distributed (improved-log) formulas apply only to their respective types of impedance and give incorrect results for the other.
However, practical accelerator components consist of both types, and it is impossible to disentangle them.
Thus, a transmission-line measurement interpretation formula is required that applies to both.
Such a formula is the (Walling-) log-formula~\cite{Walling1989}
\beq
\Zc_\parallel^\mathrm{log}=2Z_c \cdot \ln \left(\frac{S_{21}^\mathrm{REF}}{S_{21}^\mathrm{DUT}}\right)~,
\label{log-formula}
\eeq
which is obtained from Eq{.}~\eqref{improved-log-formula2} by neglecting the second term in the square bracket.
The requirement for this neglect can be conveniently expressed as
\beq
 \frac{k_\mathrm{z}^\mathrm{DUT}}{k_\mathrm{z}^\mathrm{REF}}=\frac{\Zc_{c}^\mathrm{DUT} }{\Zc_{c}^\mathrm{REF} } \approx 1~,
\label{logvalidity}
\eeq
i.e{.} the log-formula is valid if the presence of the DUT does not change the characteristic impedance significantly.
Contrary-wise, it must be invalid for a long distributed device causing a large attenuation, i.e{.} a large distributed impedance.

The systematic error of the log-formula for distributed impedances can be quantified by solving it for the logarithm and inserting into
Eq{.}~\eqref{improved-log-formula2}. The quadratic equation thereby obtained,
\beq
\Zc_\parallel^\mathrm{dist}=\Zc_\parallel^\mathrm{log}+\frac{\Zc_\parallel^{\mathrm{log}^2}}{4i\Theta_\mathrm{z}^\mathrm{REF} Z_c}~,
\eeq
has the two solutions
\beq
\Zc_\parallel^\mathrm{log}=2i\Theta_\mathrm{z}Z_c \left(-1\pm\sqrt{1+\frac{\Zc_\parallel^\mathrm{dist}}{i\Theta_\mathrm{z}^\mathrm{REF} Z_c}} \right).
\eeq
Only the positive solution is physical and gives the length independent error estimate
\beq
\frac{\Zc_\parallel^\mathrm{log}}{\Zc_\parallel^\mathrm{dist}}=
1+\frac{i}{4}\frac{\Zc_\parallel^\mathrm{dist}}{\Theta_\mathrm{z}^\mathrm{REF} Z_c}-\frac{1}{8}\left(\frac{\Zc_\parallel^\mathrm{dist}}{\Theta_\mathrm{z}^\mathrm{REF} Z_c}\right)^2
+ \cdots  \;\; ,
\label{logvaliditylumped}
\eeq
which agrees with Hahn's estimate~\cite{Hahn2000} to first order.
The systematic error of the log-formula (see Eq{.}~\eqref{log-formula}) for lumped impedance can be estimated by inserting the lumped impedance $S$-parameters (see Eq{.}~\eqref{S_lump}),
\beq
\Zc_\parallel^\mathrm{log}=-2Z_c \ln \frac{1}{1+\frac{\Zc^\mathrm{lump}}{2 Z_c}}~.
\eeq
Taylor expansion results in
\beq
\frac{\Zc_\parallel^\mathrm{log}}{\Zc^\mathrm{lump}}=1-\frac12\frac{\Zc^\mathrm{lump}}{2 Z_c}+\frac13\left(\frac{\Zc^\mathrm{lump}}{2 Z_c}\right)^2-\cdots~,
\eeq
i.e{.} the log-formula reproduces lumped impedances, for $\Zc^\mathrm{lump} \ll 2Z_c$.
Finally, one can conclude that the log formula is valid for both lumped and distributed impedance, provided the lumped part does not exceed the characteristic impedance of the REF, and the distributed part does not change the characteristic impedance significantly. Obviously, this is true for a small impedance magnitude.

\subsection{Transverse impedance}
Since the transverse impedance can be measured in a manner similar to the longitudinal one, only the different aspects are discussed.
There are two principal methods to measure the transverse impedance: the displaced-wire method and the twin-wire method.
In order to enhance the extremely small signals in the twin-wire method at low frequencies, it can be extended to the coil method, which requires a quasi-stationary interpretation.

\subsubsection{Displaced-wire method}
The displaced-wire technique is based on measuring the dipolar longitudinal impedance and using the Panofsky--Wenzel theorem to obtain the transverse impedance.
In a structure with $x$ and $y$ symmetry,
the dipolar longitudinal impedance has a quadratic dependence on the transverse offset from the centre (see e.g{.} Ref{.}~\cite{Nassibian1979}).
It can be measured in the same way as the monopolar longitudinal impedance, but with a displaced wire.
Subsequently, a parabola can be fitted on the measured results for each frequency point at different transverse positions~\cite{Metral2006}.
However, since a displaced wire measures both the driving and detuning impedance, the driving impedance in one plane can only be obtained if the detuning impedance vanishes, i.e{.} in a structure that is invariant under $90^\circ$ rotation~\cite{Burov1999}.
For rectangular structures, the detuning impedance can be cancelled by measuring the impedance in both horizontal and vertical planes and adding the two, but this yields only the sum of both driving impedances.

\subsubsection{Twin-wire method}
The setup is the same as for the longitudinal impedance, but with two symmetrically driven wires on the differential TEM mode.
The characteristic impedance (REF) for the differential TEM mode, i.e{.} the voltage between the two conductors divided by the current in one conductor, is given by (cf{.} Ref{.}~\cite{Wang2004})
\begin{align}
		 Z_c^\mathrm{dip}=\frac{Z_0}{\pi}\ln\left(\frac{d+\sqrt{d^2-a^2}}{a}\cdot\frac{b^2-d \sqrt{d^2-a^2}}{b^2+d\sqrt{d^2-a^2}}\right) ~, \label{equ_charimp_trans}
\end{align}
where $a$ is the wire radius, $b$ is outer shield radius and $2d=\Delta$ is the wire distance.
With respect to this characteristic impedance, symmetric $S$-parameters can be defined.
The symmetric $S_{21}$-parameter can be measured best with a four-port VNA, which internally converts the $4\times4$ $S$-matrix to a $2\times2$ matrix for the symmetric signals. There are also approaches to use splitters and combiners with a two-port VNA, but the limited bandwidth of those components makes the calibration (after the hybrids) an involved endeavour. For a four-port VNA the calibration plane can just be chosen before the matching section (as for the single-wire measurement) and the 18 calibration steps (open, short, match, through) can be significantly eased by using an auto-cal kit.

The twin-wire approximation provides for the ultrarelativistic dipolar transverse impedance~\cite{Nassibian1979}
\beq
\Zc_\perp(\omega)
\approx \frac{c}{\omega \Delta^2}\delta \Zc_\parallel(\omega)=\frac{c}{\omega \Delta^2} \cdot 2 Z_c^\mathrm{dip}\frac{S_{21,\mathrm{dip}}^\mathrm{REF}-S_{21,\mathrm{dip}}^\mathrm{DUT}}{S_{21,\mathrm{dip}}^\mathrm{DUT}}~,
\label{Zt_twinwire}
\eeq
where $\delta \Zc_\parallel$ is the impedance obtained by the conversion formula Eq{.}~\eqref{HPformula} for the differential mode $S_{21}$-parameter and characteristic impedance $Z_c^\mathrm{dip}$.
Since the EM-fields are mostly confined between the two wires, the de-embedded $S_{21,\mathrm{dip}}$ is very small and one does not have to distinguish between lumped and distributed transverse impedance.

\begin{figure}[t]
	\centering
   \includegraphics[angle=0,width=0.4\textwidth]{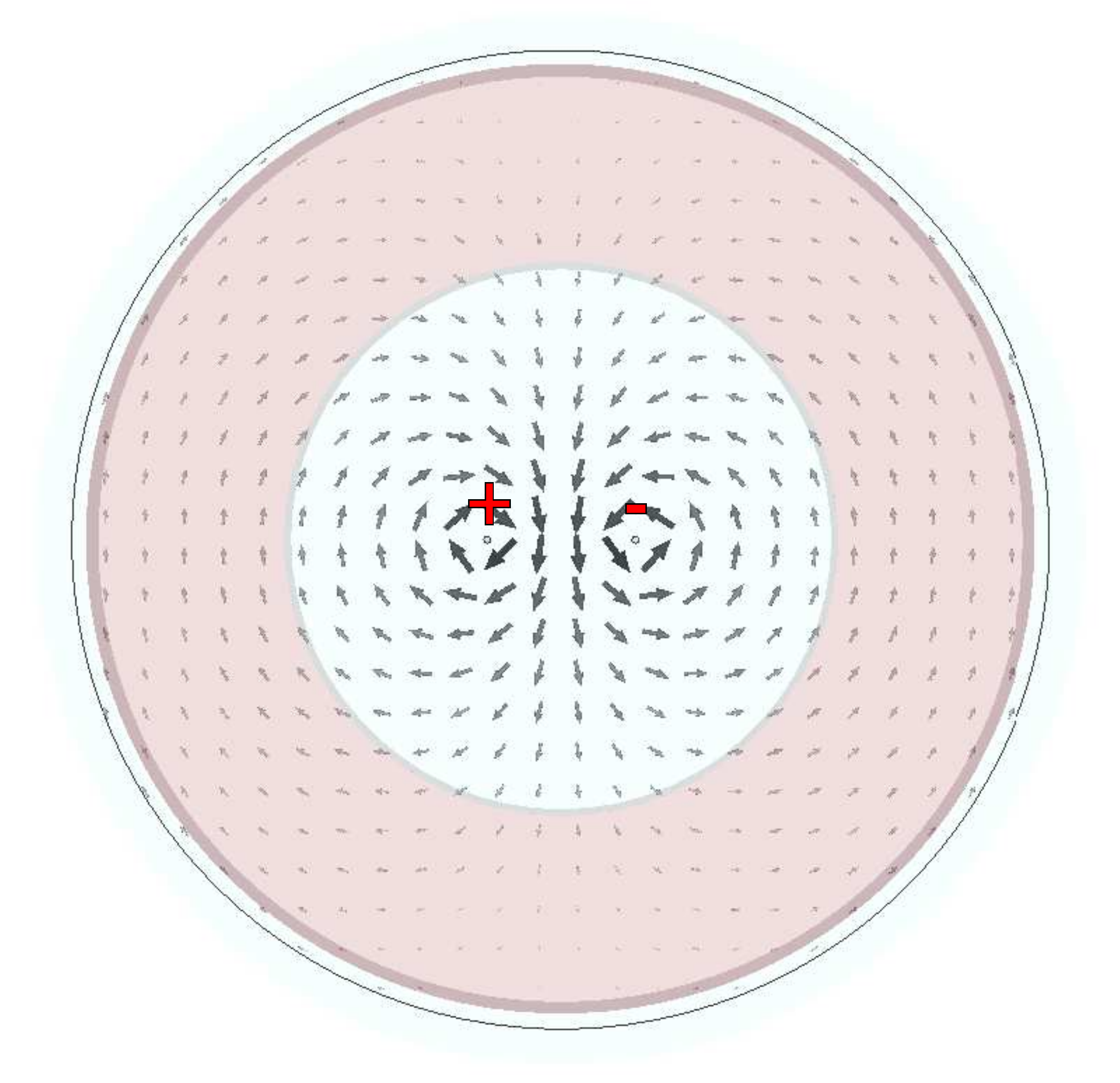}
		\includegraphics[angle=0, width=.59\textwidth]{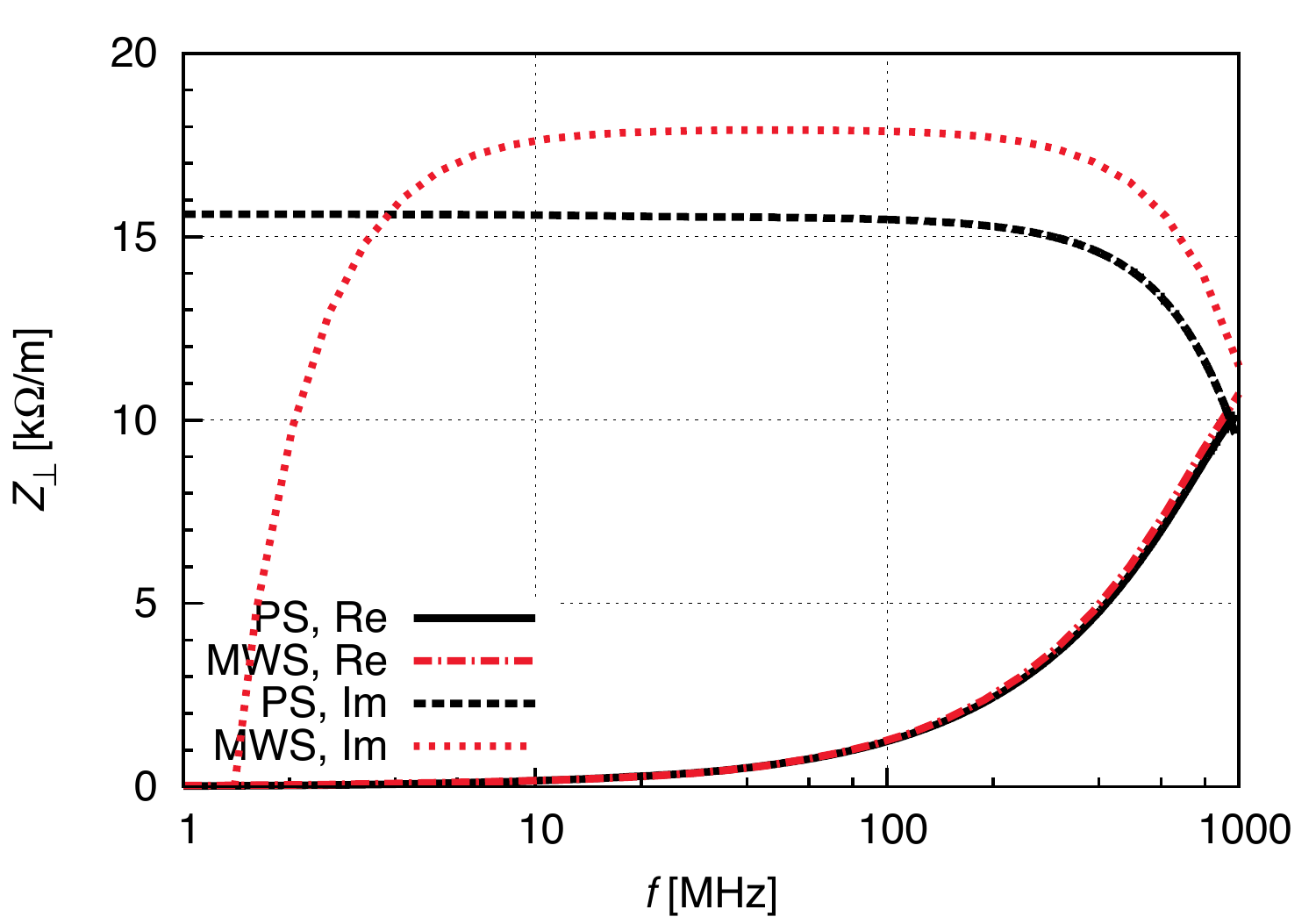}
				\caption[Magnetic field of dipole TEM eigenmode and transverse impedance from MWS and PS.]{Magnetic field of dipole TEM eigenmode obtained by the multi-pin-portmode-solver.
The wire distance is $\Delta=10$\,mm with an uncertainty of 10\% for the measurement.
The plot shows the impedance from $S$-parameter simulation (MWS) compared with wake-field simulation (PS).
}
    		\label{cst_dipolefield}
\end{figure}

\begin{figure}[h]
	\centering
		\includegraphics[angle=0, width=.49\textwidth]{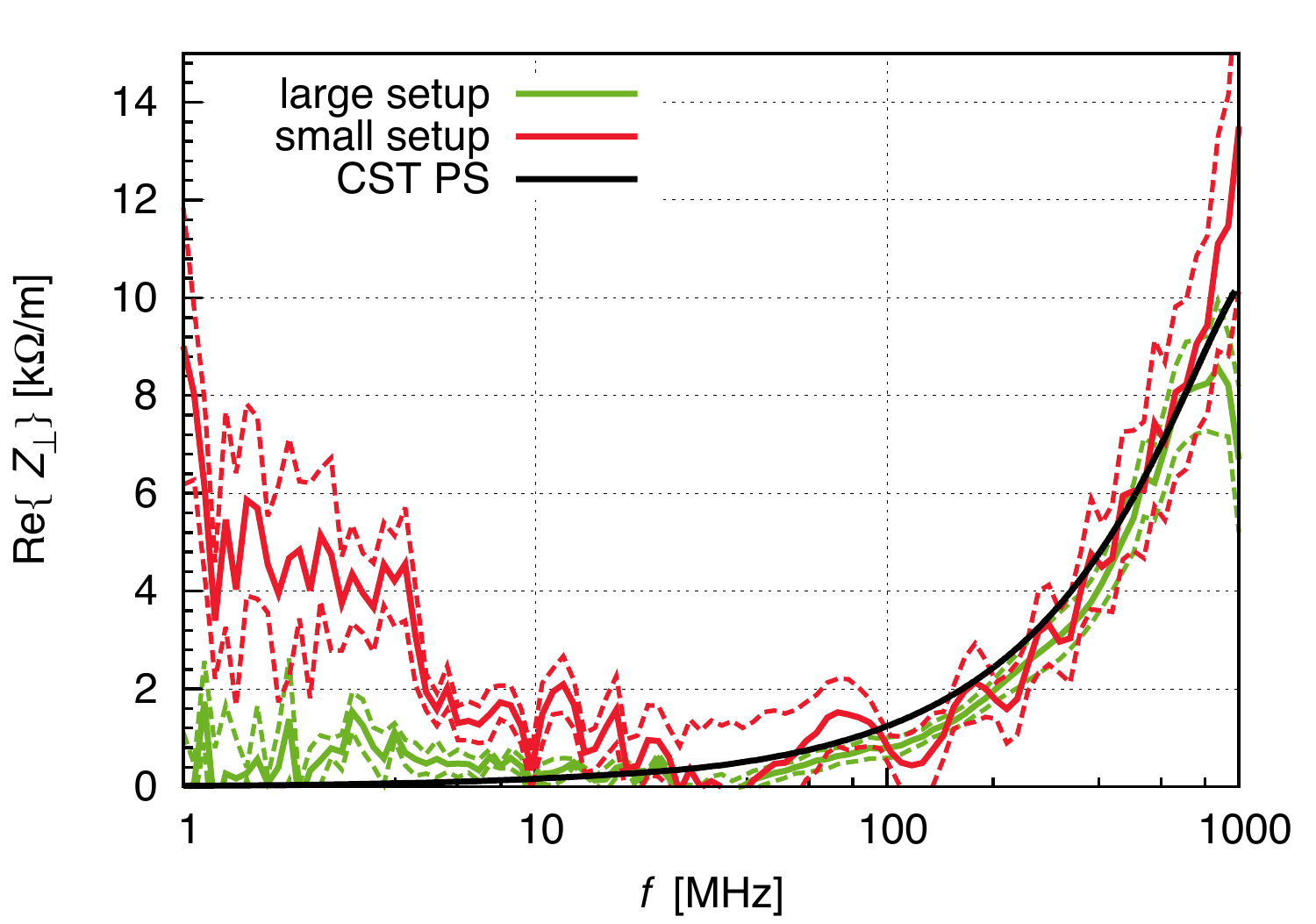}
		\includegraphics[angle=0, width=.49\textwidth]{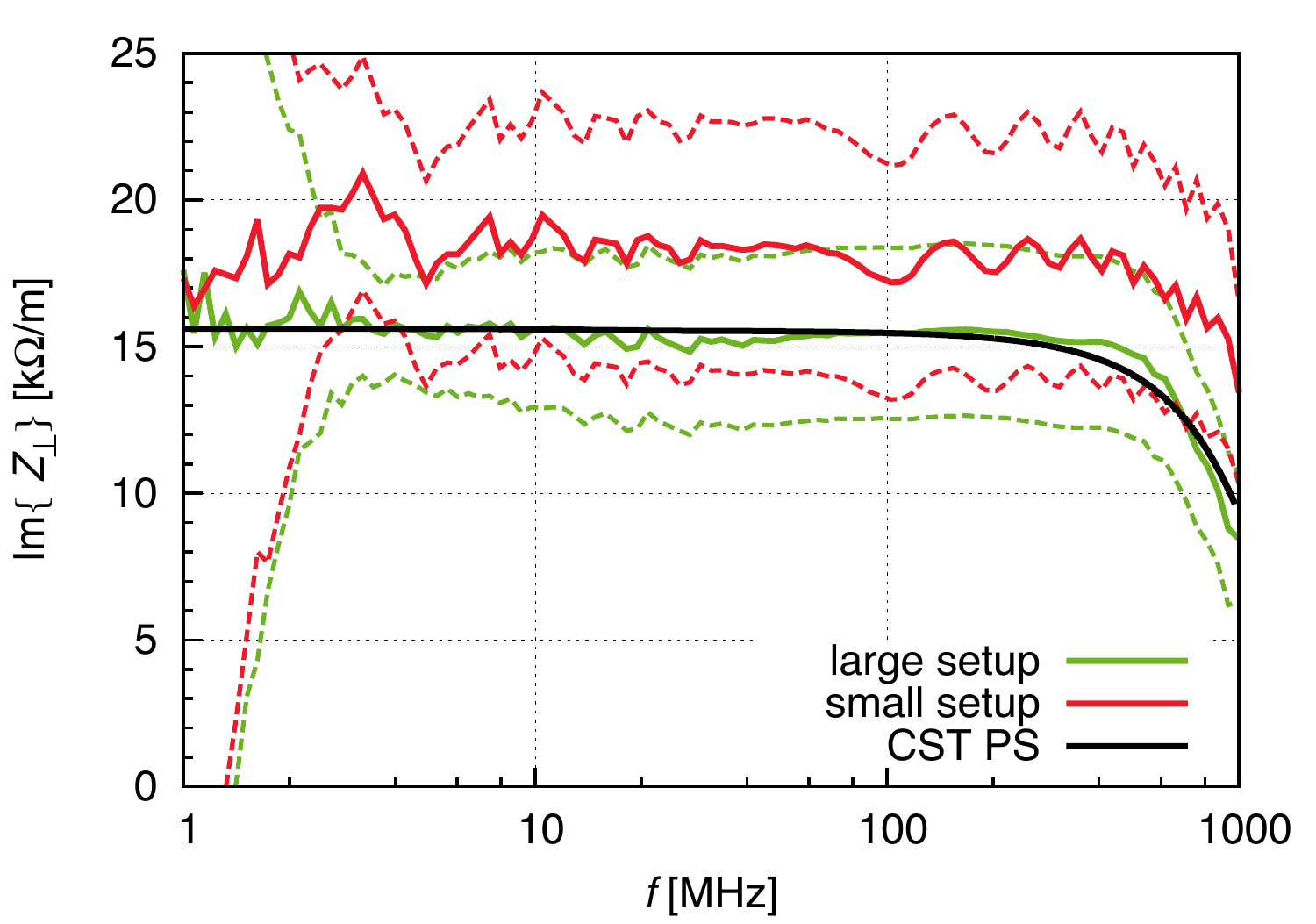}
	\caption[Measurement of the transverse impedance of the ferrite ring.]{Transverse impedance of the ferrite ring: measurement vs wake-field simulation. The dashed lines denote error bars.  }
	\label{meas_trans}
\end{figure}

A comparison of the transverse impedance from wake-field and $S$-parameter simulation for a dispersive ferrite ring is shown in Fig{.}~\ref{cst_dipolefield}.
The agreement between the two is reasonably good, except at low frequency, where the computational accuracy is insufficient.
The same is visible also for the lab measurement, as plotted in Fig{.}~\ref{meas_trans}.
Since the DUT alters the EM-fields only slightly in the twin-wire measurement, the dominating parasitic reflections are the same for DUT and REF measurements.
Thus, they are almost entirely removed by the de-embedding.
However, since the difference between DUT and REF measurement is so small, temperature drifts and noise are the main issues.
The temperature drift can be reduced by taking metal film resistors instead of carbon resistors for the matching network, which have a higher inductance but a smaller temperature coefficient.
The thereby enlarged mismatch is less critical than the temperature drift for the twin-wire measurement.
The noise can be reduced by averaging many subsequent DUT and REF measurements.

Figure~\ref{meas_trans} shows that
the result for the ferrite ring in both the large and the small measurement setup is reasonably good at medium and high frequency.
However, at low frequencies the method becomes impracticable. This can be improved by employing the coil method instead of the twin-wire method.


\subsubsection{Coil method}
\begin{figure}[h!]
	\centering
   \includegraphics[width=0.4\textwidth]{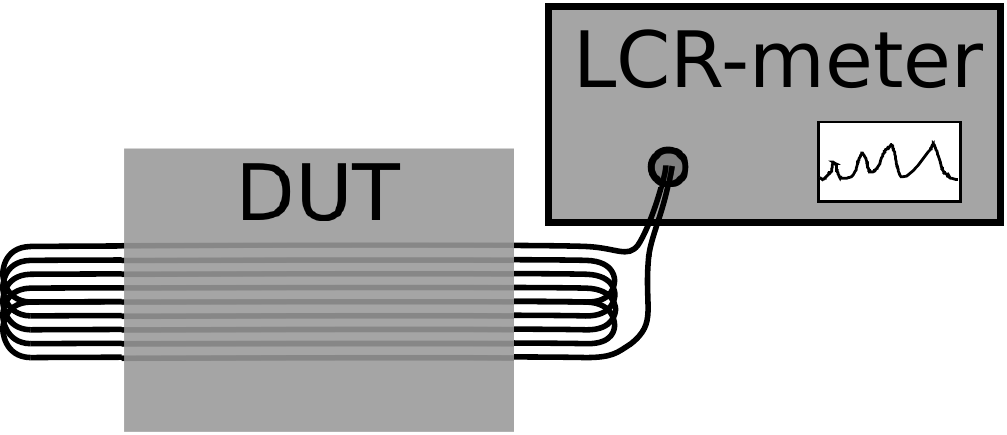}
	\hspace{1cm}
	\includegraphics[width=0.4\textwidth]{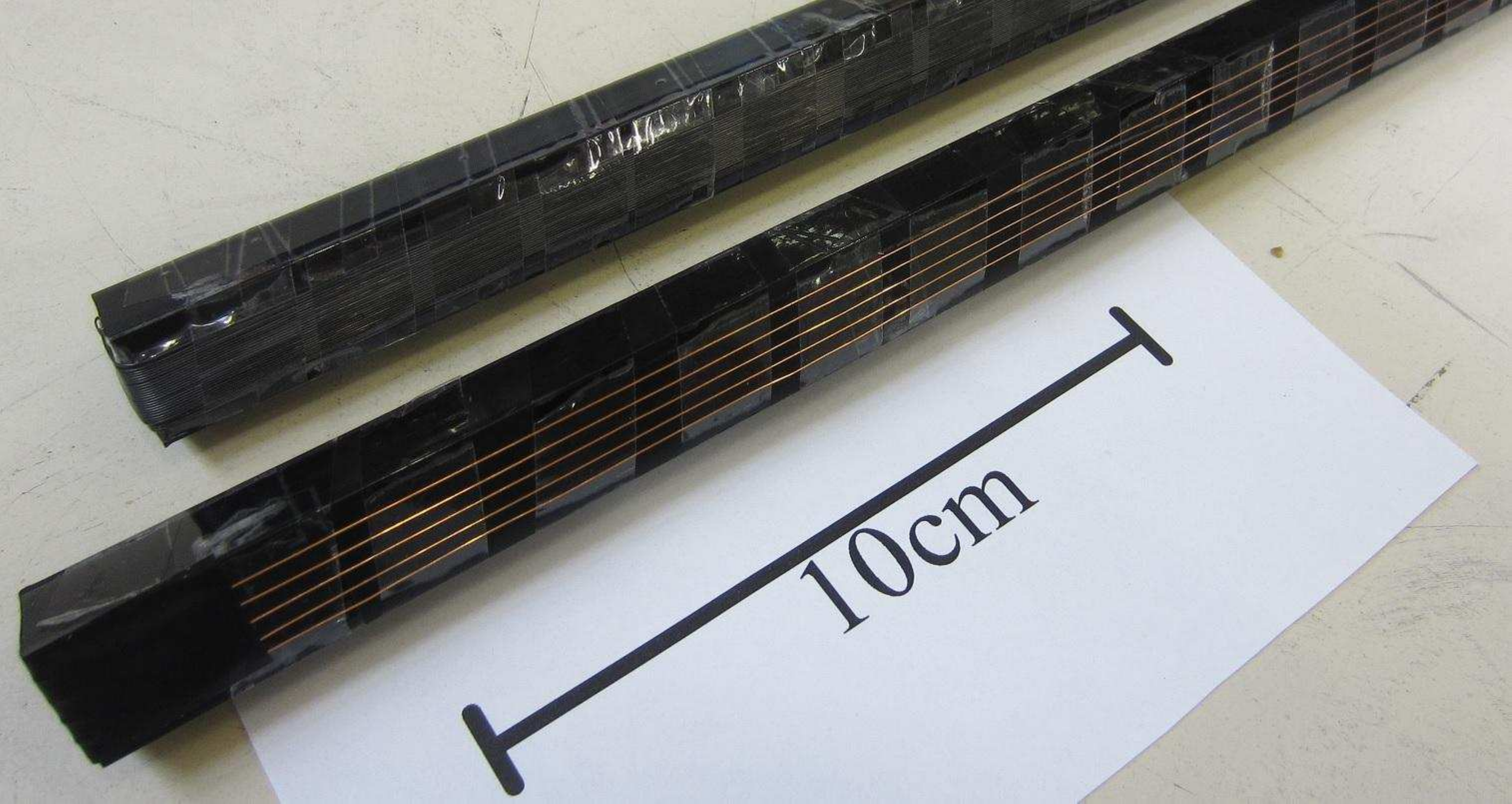}
    		\caption[Coil measurement setup. Courtesy of L.~Eidam~\cite{Niedermayer2015}.]{Coil measurement setup (left-hand side) and different coils (right-hand side). Courtesy of L. Eidam~\cite{Niedermayer2015}}
    		\label{coilsetup_sketch}
\end{figure}

In order to enhance the extremely small signals in the twin-wire method at low frequencies, the two wires can be replaced by a multi-turn coil.
Both the magnetic flux and the induced voltage are magnified by the number of turns $N$ and thus  Eq{.}~\eqref{Zt_twinwire} has to be replaced by
\begin{align}
	\Zc_\perp^\mathrm{coil}\approx\frac{c}{\omega \Delta^2 N^2} \delta \Zc~,
	\label{Zt_coil}
\end{align}
where the coil impedance difference $\delta \Zc=\Zc^\mathrm{DUT}-\Zc^\mathrm{REF}$ can be determined by a LCR-meter (we used the Agilent E4980A, 20\,Hz-2\,MHz).
The REF measurement is performed just by measuring the coil outside the DUT in free space.
Figure~\ref{coilsetup_sketch} shows the setup and two measurement coils.
The coil method has an upper frequency limit, at which the inter-turn capacitance causes a resonance which lies usually in the range of 1\,MHz.
It can be increased by taking fewer turns and increasing the turn distance.
At extremely low frequency, the accuracy is limited by the instrument noise and the temperature drift of the coil resistance.
Thus, it makes sense to use different coils: a temperature-stable one (e.g{.} constantan wire) with many turns for low frequency, and one with few turns and high conductivity (copper wire) for higher frequencies.

Since ferrites usually have very small impedance contributions at such low frequencies, the coil method is benchmarked using a steel beam pipe of 2\,mm wall thickness and 3.3\,cm radius.
\begin{figure}[h!]
	\centering
		\includegraphics[angle=-90, width=.50\textwidth]{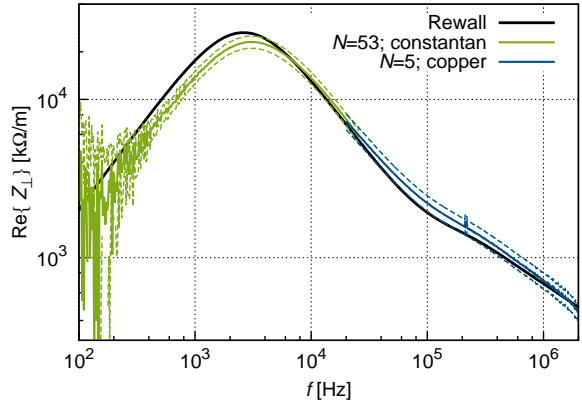}
	\caption[Coil measurement of real transverse impedance.]{Coil measurement of real transverse impedance of a tubular-beam pipe vs analytical calculation by Rewall. The dashed lines are error bars.}
	\label{coilmeasurementRe}
\end{figure}
The real part of the transverse impedance of the pipe, measured with the coils depicted in Fig{.}~\ref{coilsetup_sketch} is plotted in Fig{.}~\ref{coilmeasurementRe}.
Besides the noise at extremely low frequency, the measured real part of the impedance agrees well with the analytic prediction by Rewall~\cite{Mounet2009}.

\begin{figure}[h!]
	\centering
	\subfloat[Imaginary part of the measurement vs Rewall analytical result plus image inductance $-\text{i}5.5\times 10^4 \,\Omega /\mathrm{m}$ from Eq{.}~\eqref{ImageInductance}.]{
		\includegraphics[angle=-90, width=.48\columnwidth]{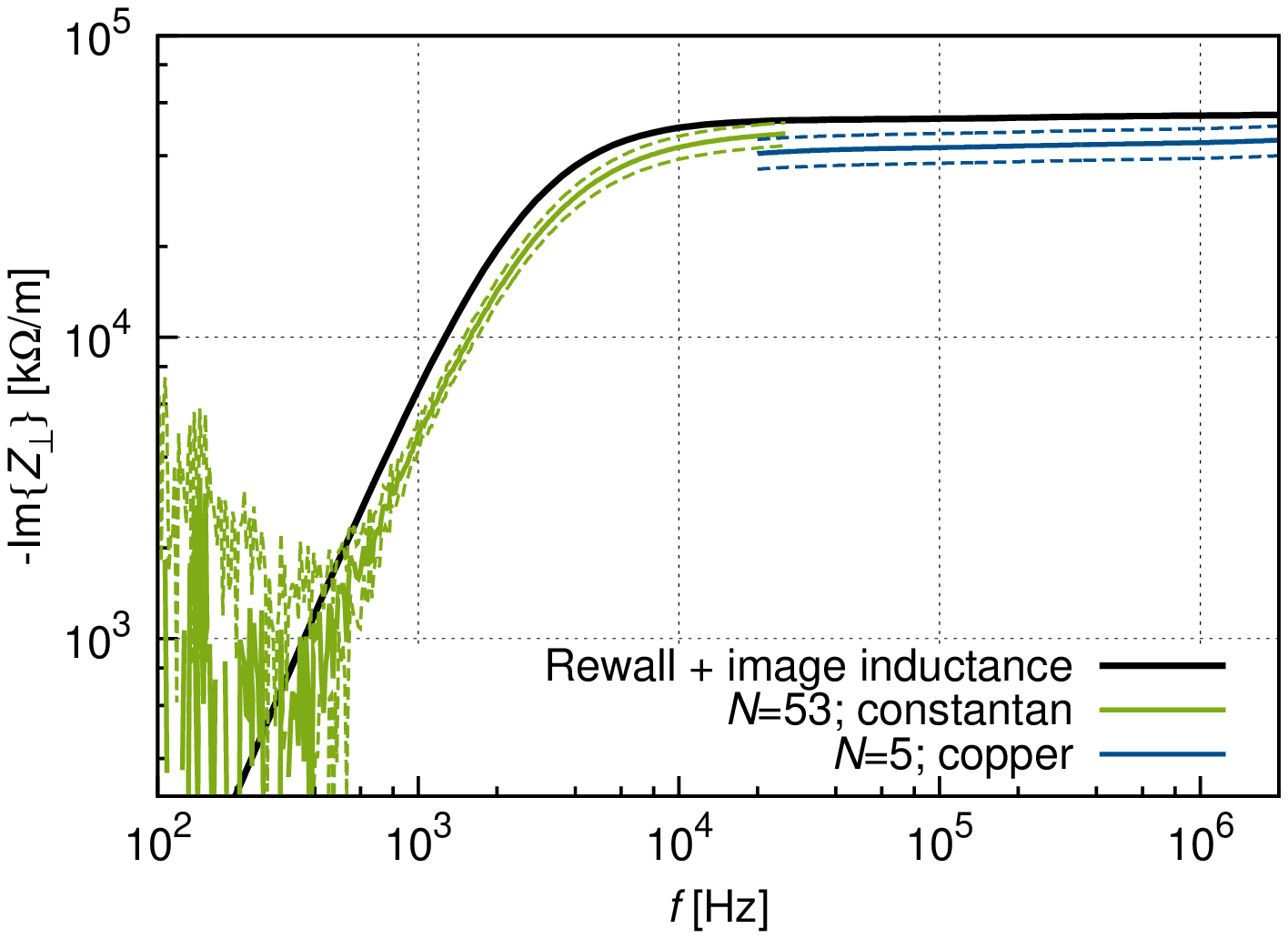}
		}
		\hfill
		\subfloat[Imaginary part of the measurement minus image inductance vs Rewall analytical result.]{
				\includegraphics[angle=-90, width=.48\textwidth]{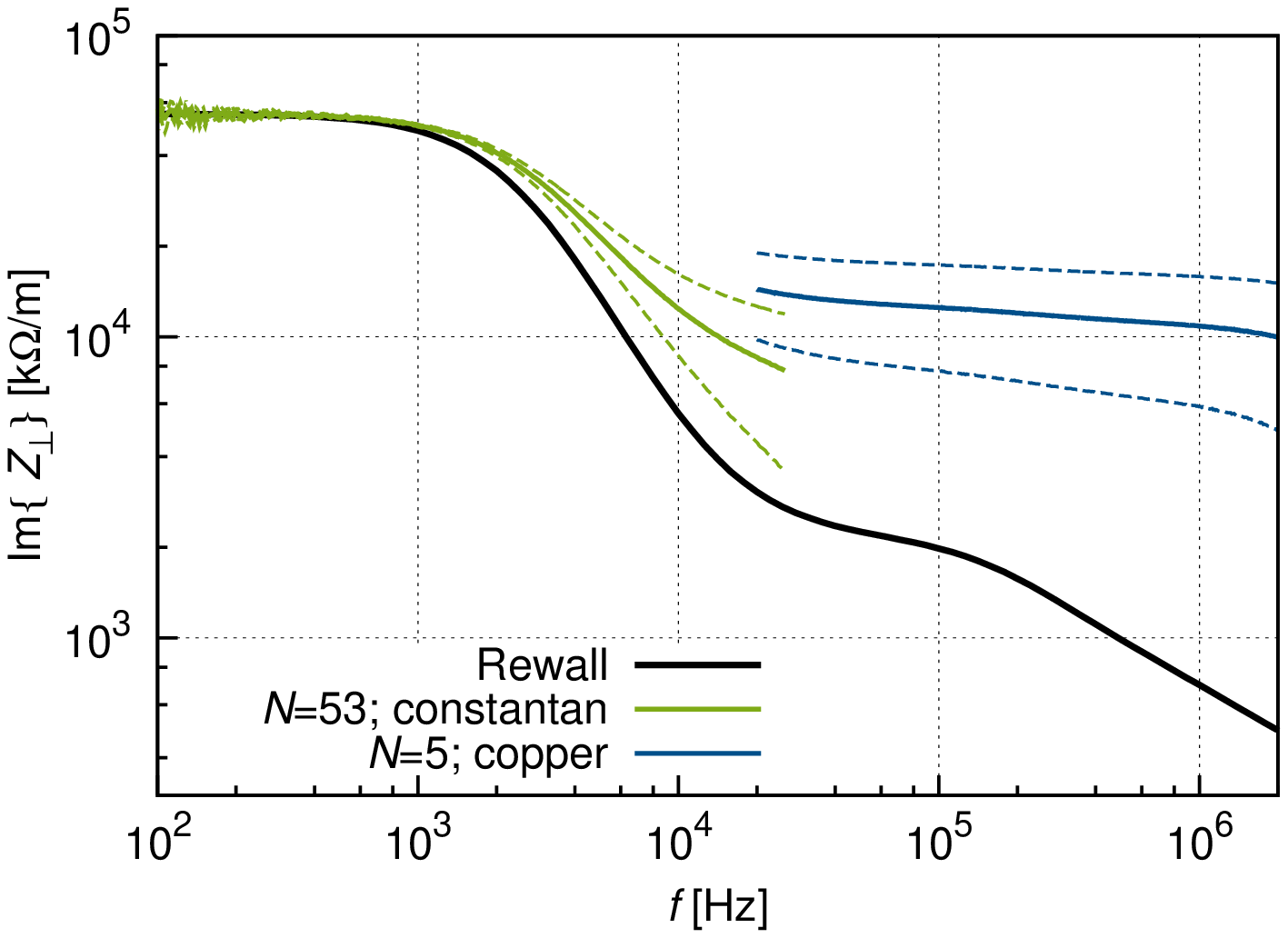}
		}
	\caption{Imaginary part from the same measurement as in Fig{.}~\ref{coilmeasurementRe}.
		Due to the high bias of the image inductance, the imaginary resistive wall impedance measurement result becomes very inaccurate.
	}
	\label{coilmeasurement2}
\end{figure}

The phase of the coil current does not depend on the longitudinal position, therefore the coil method corresponds to entirely 2D ($\partial_z=0$) source fields.
This does not represent an ultrarelativistic beam, but rather a current-only model, which is sometimes also referred to as 'radial model' and is discussed in \cite{Hahn2010,Niedermayer2012}.
The consequence of this is that the imaginary part of the impedance is superimposed by the `image inductance', i.e{.} the magnetic part of the indirect transverse space-charge impedance.
For a circular pipe of radius $b$ it is given by
\beq
\label{ImageInductance}
Z_\perp^\mathrm{image}=-\text{i}\frac{ Z_0 l}{2\pi b^2} ~. 
\eeq
The measurement result of the imaginary part of the impedance is shown in Fig{.}~\ref{coilmeasurement2}. Since the image inductance is much higher than the imaginary part of the resistive wall impedance, small relative measurement errors lead to large relative errors for the imaginary restive wall impedance.
Thus, the coil method is effective only for the determination of the real part of the transverse restive wall impedance.

\section*{Acknowledgments}
I am grateful to Lewin Eidam for providing the pictures of the bench measurements. Moreover, I would like to acknowledge fruitful discussions with the CERN ICE group.
Particularly for the bench measurements, discussions with Fritz Caspers and Manfred Wendt were enlightening.

\end{document}